\documentclass[useAMS]{emulateapj}

\usepackage{latexsym,graphicx,natbib,apjfonts}
\usepackage{amsmath}

\newcommand\simless{{\thinspace \rlap{\raise 0.5ex\hbox{$\scriptstyle  {<}$}}
    {\lower 0.3ex\hbox{$\scriptstyle  {\sim}$}} \thinspace }}  
\newcommand\simgreat{{\thinspace \rlap{\raise 0.5ex\hbox{$\scriptstyle  {>}$}}
    {\lower 0.3ex\hbox{$\scriptstyle  {\sim}$}} \thinspace }}  

\newcommand\msun{\, \rm M_\odot} 
\newcommand\rsun{\, \rm R_\odot} 
\newcommand\kms{{\, \rm km\,s^{-1}}}
\newcommand\yr{{\, \rm yr}}
\newcommand\myr{{\, \rm Myr}}
\newcommand\mpc{{\, \rm mpc}}
\newcommand\pc{{\, \rm pc}}
\newcommand\mbh{M_{\rm SMBH}}
\newcommand\mim{M_{\rm IMBH}}
\newcommand\mstar{m_{\rm *}}
\newcommand\mtot{M_{\rm SMBH} + M_{\rm IMBH}}
\newcommand\abh{a_{\rm BH}}
\newcommand\tdf{T_{\rm DF}}
\newcommand\tpr{T_{\rm PR}}
\newcommand\tgw{T_{\rm GW}}
\newcommand\ainn{a_{\rm inn}}
\newcommand\aout{a_{\rm out}}
\newcommand\einn{e_{\rm inn}}
\newcommand\eout{e_{\rm out}}
\newcommand\Pinn{P_{\rm inn}}
\newcommand\Pout{P_{\rm out}}
\newcommand\mas{\, \rm mas}
             
\shorttitle{Perturbations on Stellar Orbits in the Galactic Center}
\shortauthors{Gualandris and Merritt}

\begin{document}

\title{Perturbations of Intermediate-mass Black Holes on Stellar
  Orbits in the Galactic Center}

\author{Alessia Gualandris and David Merritt}

\affil{Department of Physics and Center for Computational Relativity
  and Gravitation, Rochester Institute of Technology, 78 Lomb Memorial
  Drive, Rochester, NY}

\email{alessiag,merritt@astro.rit.edu}

\begin{abstract}
We study the short- and long-term effects of an intermediate mass
black hole (IMBH) on the orbits of stars bound to the supermassive
black hole (SMBH) at the center of the Milky Way.  A regularized
$N$-body code including post-Newtonian terms is used to carry out
direct integrations of 19 stars in the S-star cluster for $10\myr$.
The mass of the IMBH is assigned one of four values from $400\msun$ to
$4000\msun$, and its initial semi-major axis with respect to the SMBH
is varied from $0.3-30\mpc$, bracketing the radii at which inspiral of
the IMBH is expected to stall.  We consider two values for the
eccentricity of the IMBH/SMBH binary, $e=(0,0.7)$, and 12 values for
the orientation of the binary's plane.  Changes at the level of $\sim
1\%$ in the orbital elements of the S-stars could occur in just a few
years if the IMBH is sufficiently massive.  On time scales of $1\myr$
or longer, the IMBH efficiently randomizes the eccentricities and
orbital inclinations of the S-stars. Kozai oscillations are observed
when the IMBH lies well outside the orbits of the stars. Perturbations
from the IMBH can eject stars from the cluster, producing
hypervelocity stars, and can also scatter stars into the SMBH; stars
with high initial eccentricities are most likely to be affected in
both cases.  The distribution of S-star orbital elements is
significantly altered from its currently-observed form by IMBHs with
masses greater than $\sim 10^3\msun$ if the IMBH/SMBH semi-major axis
lies in the range $3-10\mpc$.  We use these results to further
constrain the allowed parameters of an IMBH/SMBH binary at the
Galactic center.
\end{abstract}


\keywords{Galaxy:center - stellar dynamics}

\section{Introduction}
Near-infrared imaging and spectroscopy of the Galactic center indicate
the presence of over 100 young massive stars in the inner parsec of
the Milky Way. These belong to two distinct populations: (1) Outside
the central arcsecond, a group of roughly 40 stars, mostly O and
Wolf-Rayet stars with an estimated age of $6\myr$, move on
approximately circular orbits in a thin and coherent stellar disk
\citep{paumard06} (2) Inside the central arcsecond, a group of 20
B-type stars (the S-stars) move on eccentric and randomly oriented
orbits around the supermassive black hole (SMBH)
\citep{ghez03,eisen05,gill09}.

The existence of young massive stars so close to the SMBH is puzzling
since tidal stresses from the SMBH are likely to suppress star
formation in this region. If the S-stars formed at larger distances
from Sgr A$^*$ and were then transported inward, the migration would
need to occur on a timescale of only a few Myr, given the apparent age
of the stars. This constitutes the so-called {\it paradox of youth}.

Several scenarios have been proposed for the origin of the young stars
in the Galactic center, including (i) rejuvenation of older stars due
to physical collisions and/or tidal stripping in the vicinity of the
SMBH \citep{lee96, ghez03}, (ii) tidal disruption of binary stars on
low angular momentum orbits \citep{gq03, gl06}, (iii) in-situ
formation from a fragmenting gas disk formed from an infalling
molecular cloud \citep{l07, bonnrice08}, (iv) formation in a star
cluster outside the central parsec and transportation by the
inspiralling cluster \citep{gerhard01}.

The spectra of the observed S-stars seem to indicate that they are
ordinary B-type main-sequence stars, with no apparent evidence for an
exotic history \citep{eisen05,figer08}. This would seem to exclude the
rejuvenation scenario.

The binary disruption scenario postulates that the S-stars were once
in binary systems on plunging orbits that were tidally disrupted by
the SMBH, ejecting one star and leaving the other in an extremely
eccentric orbit around the SMBH \citep{gq03}. This model requires an
ad hoc reservoir of new binaries at large radii, a mechanism to
scatter the stars onto plunging orbits (e.g. ``massive perturbers'';
\citet{perets07}), and an efficient mechanism to thermalize the
eccentricities within the lifetime of the stars (e.g. ``resonant
relaxation'' from stellar mass black holes; \citet{perets09}).
Modified binary disruption scenarios in which the S-stars were born in
a stellar disk have also been proposed \citep{ma09, lock09}.

The in-situ formation model assumes that an infalling molecular cloud
interacts with the SMBH and forms a disk which then fragments and
forms stars. This model successfully reproduces the observed
properties of the disk stars \citep{bonnrice08}, including mild
eccentricities and a top-heavy mass function.  It however cannot
account for the large eccentricities and random inclinations of the
orbits of the S-stars.

\citet{mgm09} showed that these dynamical properties can be naturally
explained by interaction of the stars with an intermediate mass black
hole (IMBH) in the context of the inspiralling star cluster
scenario. This model postulates that the S-stars were born in a star
cluster at large enough distance from the SMBH that a giant molecular
cloud could collapse and fragment. The new cluster then migrated
inward due to dynamical friction, depositing stars along the way as
the tidal force from the SMBH stripped them from the outer regions of
the cluster \citep{gerhard01}.  If the cluster has by then formed an
IMBH at its center, e.g. via the runaway merging scenario
\citep{spz02}, the cluster can remain bound and spiral inward for much
longer before releasing all its component stars \citep{hm03}, reaching
distances of a few tens of milliparsecs before evolution of the
IMBH/SMBH binary stalls \citep{baum06,mme07,lock08}.  Interactions
with the same IMBH that brought the stars close to the SMBH can then
randomize their orbital inclinations and thermalize their eccentricity
distribution on timescales of $\simless 1\myr$ \citep{mgm09}.  The
IMBH can also capture stars ejected from the cluster into a resonance
and bring them close to the SMBH \citep{fujii09}.

Whether or not the inspiralling star cluster model is correct, this
work raises intriguing questions about the observable consequences of
an IMBH that is currently imbedded in the S-star cluster.  Are
perturbations from an IMBH detectable on short (year-long) time
scales, and if so, which of the S-stars are most likely to be
affected?  How stable is the cluster on time scales of order stellar
lifetimes?  Would an appreciable fraction of the stars be ejected from
the Galactic center, or scattered into the SMBH, by the IMBH?  What
constraints can be placed on the physical parameters (mass, orbital
elements) of a putative IMBH/SMBH binary given the need to conserve
the observed distribution of orbital elements of the S-stars?

To address these questions, we have carried out a large set of
high-accuracy $N$-body simulations, taking as our initial conditions
the current positions and velocities of the $\sim 20$ S-stars with
well determined orbits, plus an additional massive particle
representing the IMBH.  In \S 2 we describe the initial conditions and
the $N$-body methods.  \S3 discusses the short (few years) time scale
effects on the observed stellar orbits.  The longer term ($\myr$)
behavior of the S-star cluster is described in \S4.  \S 5 summarizes
our results and draws conclusions about the allowed parameter space
for an IMBH/SMBH binary at the Galactic center.

\section{Initial models and numerical methods}

Our $N$-body models include a SMBH, an IMBH and a cluster of 19
S-stars.  \citet{gill09} provide Keplerian elements for 28 stars, of
which one (S111) appears to be unbound as a result of its large radial
velocity, six (S66, S67, S83, S87, S96, S97) likely belong to the
clockwise stellar disk \citep{genzel03, paumard06} and one (S71) has a
very long orbital period.  The remaining stars have well defined or at
least reliable orbits and exhibit essentially random orbital
orientations.  We determined positions and velocities at year 2008 AD
for the 19 remaining stars from the classical elements\footnote{The
  orbital elements are defined as in \citet{eisen05} and
  \citet{paumard06}: $a$ = semi-major axis; $P$ = orbital period; $e$
  = eccentricity; $i$ = inclination of the ascending node, i.e. angle
  between the orbital plane and the plane tangential to the celestial
  sphere ($0^{\circ} < i < 90^{\circ}$ for direct counterclockwise
  projected rotation, $90^{\circ} < i < 180^{\circ}$ for retrograde
  clockwise projected rotation); $\Omega$ = position angle of the
  ascending node; $\omega$ = longitude of periastron, i.e. the angle
  between the radius vector of the ascending node and that of the
  periastron counted from the node in the direction of the orbital
  motion.} given in Table\,\ref{tab:sstars}.
\begin{table}
\begin{center}
\caption{S-star orbital elements}
\label{tab:sstars}
\begin{tabular}{ccccccc}
\hline 
star  & $P_{\rm orb}$ & $a$ & $e$ & $i$ & $\Omega$ & $\omega$\\
      & (yr) & (mpc) & & (deg) & (deg) & (deg) \\
\hline 
S1  &  128.0   &  19.462    &  0.483  &   120.1 &  341.3 &  116.9 \\ 
S2  &   15.8   &   4.829    &  0.880  &   134.5 &  225.8 &   63.8\\
S4  &   58.2   &  11.523    &  0.407  &    77.8 &  257.9 &  316.9\\
S5  &  210.0   &  27.072    &  0.672  &   110.2 &  147.0 &  270.6\\
S6  &  119.0   &  18.601    &  0.898  &    86.5 &   84.0 &  126.2\\
S8  &   95.6   &  16.038    &  0.816  &    74.6 &  315.7 &  344.4\\
S9  &   62.3   &  12.054    &  0.867  &    80.4 &  145.7 &  229.7\\
S12 &   60.1   &  11.763    &  0.899  &    32.8 &  236.9 &  311.2\\
S13 &   56.3   &  11.263    &  0.471  &    27.8 &   72.0 &  247.5\\
S14 &   45.2   &   9.727    &  0.962  &    99.4 &  227.9 &  339.0\\
S17 &   62.0   &  12.019    &  0.370  &    96.3 &  187.2 &  318.7\\
S18 &   47.9   &  10.123    &  0.748  &   114.4 &  216.8 &  150.5\\
S19 &  323.0   &  36.140    &  0.863  &    74.6 &  343.7 &  157.0\\
S21 &   35.0   &   8.203    &  0.784  &    55.3 &  254.8 &  178.0\\
S24 &  444.0   &  44.677    &  0.941  &   106.6 &    3.0 &  290.1\\
S27 &   95.9   &  16.072    &  0.963  &    93.3 &  192.1 &  312.9\\
S29 &   85.1   &  14.839    &  0.936  &   127.0 &  157.3 &  346.3\\
S31 &   64.2   &  12.303    &  0.699  &   115.1 &  138.9 &  343.4\\
S33 &   95.0   &  15.967    &  0.731  &    44.5 &   82.7 &  326.9\\
\hline
\end{tabular}
\end{center}
\end{table}
The assumed SMBH mass is $\mbh=4.0\times10^6\msun$ \citep{gill09}. The
masses of the S-stars were set to $10\msun$ \citep[e.g.][]{eisen05}
except for star S-2 for which a value of $20\msun$ was adopted
\citep{martins08}.

We considered four values of the IMBH mass, $\mim
=(400,1000,2000,4000)\msun$, or $q=\mbh/\mim=(1, 2.5, 5, 10)\times
10^{-4}$; five values of the initial semi-major axis, $a = (0.3, 1, 3,
10, 30)\mpc$; two values of the eccentricity, $e = (0, 0.7)$; and
twelve orientations of the IMBH/SMBH orbital plane, for a total of 480
initial models.  The parameters of the simulations are summarized in
Table\,\ref{tab:ic}.
\begin{table}
\begin{center}
\caption{Parameters of the simulations}
\label{tab:ic}
\begin{tabular}{cccccc}
\hline
 Run     &  Run & $q$ & $\mim$ & $a$ & $P$  \\
 e = 0.0   & e = 0.7 & & ($\msun$) & (mpc) & (yr) \\
\hline
  1-12    &  241-252    & $1.0\times10^{-4}$ & 400 & 0.3 &   0.243\\
 13-24    &  253-264    & $1.0\times10^{-4}$ & 400 &   1 &   1.481\\
 25-36    &  265-276    & $1.0\times10^{-4}$ & 400 &   3 &   7.696\\
 37-48    &  277-288    & $1.0\times10^{-4}$ & 400 &  10 &  46.837\\
 49-60    &  289-300    & $1.0\times10^{-4}$ & 400 &  30 & 243.373\\
\hline
 61-72    & 301-312    & $2.5\times10^{-4}$ & 1000 & 0.3 &   0.243\\
 73-84    & 313-324    & $2.5\times10^{-4}$ & 1000 &   1 &   1.481\\
 85-96    & 325-336    & $2.5\times10^{-4}$ & 1000 &   3 &   7.695\\
 97-108   & 337-348    & $2.5\times10^{-4}$ & 1000 &  10 &  46.834\\
109-120   & 349-360    & $2.5\times10^{-4}$ & 1000 &  30 & 243.355\\
\hline
121-132   & 361-372    & $5.0\times10^{-4}$ & 2000 & 0.3 &   0.243\\
133-144   & 373-384    & $5.0\times10^{-4}$ & 2000 &   1 &   1.481\\
145-156   & 385-396    & $5.0\times10^{-4}$ & 2000 &   3 &   7.695\\
157-168   & 397-408    & $5.0\times10^{-4}$ & 2000 &  10 &  46.828\\
169-180   & 409-420    & $5.0\times10^{-4}$ & 2000 &  30 & 243.324\\
\hline
181-192   & 421-432    & $1.0\times10^{-3}$ & 4000 & 0.3 &   0.243\\
193-204   & 433-444    & $1.0\times10^{-3}$ & 4000 &   1 &   1.480\\
205-216   & 445-456    & $1.0\times10^{-3}$ & 4000 &   3 &   7.692\\
217-228   & 457-468    & $1.0\times10^{-3}$ & 4000 &  10 &  46.816\\
229-240   & 469-480    & $1.0\times10^{-3}$ & 4000 &  30 & 243.263\\
\hline
\end{tabular}
\end{center}
\end{table}

\begin{figure}
  \begin{center}
    \includegraphics[width=6.5cm,angle=270]{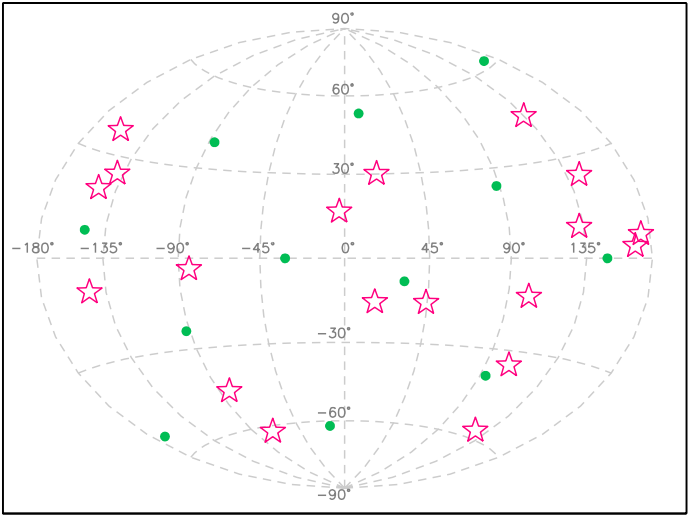}
  \end{center}
  \caption{Directions of the angular momentum vector of the 19 S-stars
    (stars) and of the 12 possible orientations  of the IMBH/SMBH orbit
    (circles). In this all-sky map, the vertical dimension refers to
    the inclination $i$ of the orbit and the horizontal dimension to
    the longitude of the ascending node $\Omega$, as defined in
    \citet{eisen05} and \citet{gill09}. The Galactic plane lies at
    $i=0^{\circ}$, $\Omega=-31.4^{\circ}$.}
  \label{fig:map}
\end{figure}

The orientations of the massive binary were not selected randomly but
rather were distributed uniformly on a grid.  The surface of a sphere
can be uniquely tesselated by means of 12 regular pentagons, the
centers of which form a regular dodecahedron inscribed in the sphere.
We determined the centers of the pentagons and transformed the
coordinates of the points into inclination $i$ and longitude of
ascending node $\Omega$ of the IMBH/SMBH orbit.  We then rotated the
points by an arbitrary angle to avoid chance superpositions with the
S-star orbital inclinations.  The resulting distribution of angular
momentum vectors on the plane of the sky is shown in
Figure~\ref{fig:map}; the values of the 12 angles are given in
Table~\ref{tab:angles}.

\begin{table}
\begin{center}
\caption{Grid of IMBH/SMBH orbital inclinations}
\label{tab:angles}
\begin{tabular}{ccc}
\hline 
number  & $i$ (deg) & $\Omega$ (deg)\\
\hline 
     1 &    90.0   &   31.49 \\
     2 &   129.21  &   86.17 \\
     3 &    50.79  &  266.17 \\
     4 &    90.0   &  211.49 \\
     5 &    27.97  &   14.14 \\
     6 &    81.95  &  328.25 \\
     7 &    98.05  &  148.23 \\
     8 &   152.03  &  194.14 \\
     9 &   66.07   &   92.05 \\
    10 &   37.09   &  169.12 \\
    11 &   142.91  &  349.12 \\
    12 &   113.93  &  272.05 \\
\hline
\end{tabular}
\end{center}
\end{table}

We followed the evolution of each 21-body system for a time
corresponding to $10\myr$ using the AR-CHAIN code \citep{mm08}, a
recent implementation of the algorithmic regularization method that is
able to reproduce the motion of tight binaries for long periods of
time with extremely high precision.  The code combines the use of the
chain structure, introduced originally by \cite{ma93}, with a new
time-transformation to avoid singularities and achieve high precision
for arbitrary mass ratios. 
Integrations were carried out on two computer clusters at the Center
for Computational Relativity and Gravitation at the Rochester
Institute of Technology: {\tt gravitySimulator} \citep{harfst07} and
{\tt newHorizons}\footnote{http://ccrg.rit.edu/facilities}, and on one
computer cluster hosted by the Research Computing group at the same
institution.

\begin{figure}
  \begin{center}
    \includegraphics[width=8.5cm]{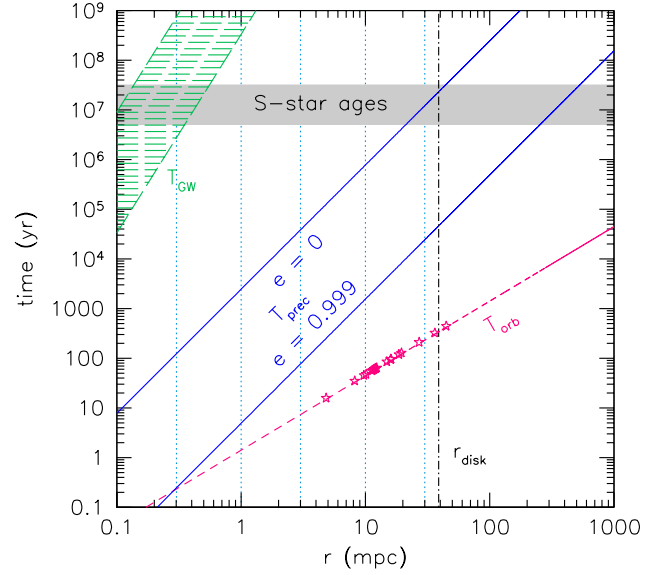}
  \end{center}
  \caption{Relevant timescales for the S-stars and the IMBH as a
    function of distance from the SMBH: $T_{\rm orb}$ orbital period
    (dashed line); star symbols are the S-stars, $T_{\rm prec}$
    relativistic precession timescale (solid lines) for eccentricities
    $e=0$ and $e=0.999$, $\tgw$ gravitational wave emission timescale
    for $\mim =400-4000\msun$ and $e=0-0.7$ (dashed region).  The
    vertical dotted lines represent the adopted values for the black
    hole binary initial separation. The filled grey region indicates
    the estimated ages of the S-stars. The vertical dotted-dashed line
    represents the outer edge of the disk of massive stars.}
  \label{fig:times}
\end{figure}

The AR-CHAIN code also includes
post-Newtonian corrections to the accelerations up to order
PN2.5. This is necessary due to the fact that the general relativistic
precession timescale (accounted for by the PN1 and PN2 terms)
\begin{eqnarray}
  \tpr & = & \frac{2 \pi c^2 \left(1-e^2\right) a^{5/2}}{3 \left(G \mbh \right)^{3/2}} \nonumber \\ 
       & \simeq &  8.2\times10^5 {\rm yr} \left(\frac{a}{10\mpc}\right)^{5/2} \left(\frac{4\times10^6\msun}{\mbh}\right)^{3/2}
\left(1-e^2\right)
\end{eqnarray}
is generally much shorter than the integration time (see also
Fig.\,\ref{fig:times}). The energy loss due to emission of
gravitational waves (accounted for by the PN2.5 term) occurs on a
timescale
\begin{eqnarray}
\label{eq:tgw}
  \tgw & = & \frac{5}{256 F(e)} \frac{c^5}{G^3} \frac{a^4}{\mu
    \left(\mtot\right)^2} \nonumber\\ & \simeq &
  \frac{3.6\times10^{10} \yr}{F(e)} \left(\frac{a}{10\mpc}\right)^4
  \left(\frac{4\times10^6\msun}{\mbh}\right)
  \left(\frac{10^3\msun}{\mim}\right)\times \nonumber\\ & & \times
  \left(\frac{4\times10^6\msun}{\mtot}\right)
\end{eqnarray}
where
\begin{equation}
  F(e) = \left(1-e^2\right)^{-7/2} \left(1 + \frac{73}{24}e^2 + \frac{37}{96} e^4\right)
\end{equation}
and
\begin{equation}
  \mu = \frac{\mbh * \mim}{\mtot} \approx q \mim\,.
\end{equation}
This is typically $\simgreat 10\myr$ (see Fig.\,\ref{fig:times}),
implying that the orbit of the IMBH does not evolve appreciably over
the integration time. The only exception is the set of runs with
$q=10^{-3}$, $a=0.3\mpc$ and $e=0.7$, for which $\tgw \simless 3\myr$.

Some relevant timescales for the S-stars and the IMBH are shown in
Figure\,\ref{fig:times}.  One timescale that is not illustrated there is
that associated with dynamical friction of the IMBH against the
background stars.  Assuming a circular orbit for the IMBH, the
frictional drag is given by \citep[Eq.7-18]{bt87}
\begin{equation}
F = - \frac{4\pi\,\rm ln\Lambda G^2 \rho(r)\, \mim^2}{v_c^2} \left(\rm erf(X) - \frac{2X}{\sqrt{\pi}}e^{-X^2}\right)
\end{equation}
where $\rho(r)$ is the background stellar density, $\rm ln\Lambda$ the
Coulomb logarithm and $X = v_c / \sqrt{2} \sigma$ is the ratio of the
IMBH velocity to the one-dimensional velocity dispersion $\sigma$.
The rate of change of the angular momentum $L = r\,v_c =
\sqrt{G\mbh\,r}$ can be written as $dL/dt = F\,r / \mim$.  Assuming
$X=1$, this implies
\begin{equation}
r^{-5/2} \frac{dr}{dt} = - 3.4 \pi\,\rm ln\Lambda \sqrt{G} \frac{\mim}{\mbh^{3/2}} \rho(r) 
\end{equation}
The dynamical friction timescale is then
\begin{eqnarray}
&&\tdf = \left| \frac{1}{r} \frac{dr}{dt} \right|^{-1}  =  \frac{1}{3.4\pi \,\rm ln\Lambda \sqrt{G}} \frac{\mbh^{3/2}}{\mim} \frac{r^{-3/2}}{\rho(r)} \\
&& \simeq  2\times10^7 \yr \left(\frac{\mbh}{4\times10^6\msun}\right)^{3/2} \left(\frac{10^3\msun}{\mim}\right) \left(\frac{10^5 \msun \pc^{-3}}{\rho}\right) \left(\frac{r}{\pc}\right)^{-3/2} \nonumber\\
\end{eqnarray}
having taken $\rm ln\Lambda = 6.6$ \citep{spinnato03}.  

Since our $N$-body models include only a fraction of the stellar mass 
in this region, the IMBH experiences essentially no dynamical friction.  
This may, or may not, be an accurate representation of what would
actually happen at the Galactic center.
Recent observations \citep{bu09,do09} raise
questions about the existence of a density cusp in the observed
stellar population within the central parsec:
Number counts of the late-type (old) stars suggest a density that 
{\it drops} inside of $\sim 0.5$ pc.
Proper-motion data \citep{schodel09} are also consistent with
a mass density that falls toward SgrA$^*$.
Even if a density cusp were present initially, an inspiralling IMBH 
tends to displace the background stars and create a low-density core
on a scale of $\sim 0.1$ pc \citep{baum06}.
The absence of dynamical friction in our simulations may therefore
be a reasonable approximation.

By ignoring the possibility of a density cusp, we also exclude
the other dynamical effects that would accompany a high  density
of background stars, including cusp-induced precession of the orbits;
damping of Kozai oscillations; and ejection of S-stars by close 
interactions with stellar remnants.

\section{Results: Short-term evolution}

\begin{figure*}
  \begin{center}
    \includegraphics[width=13.cm,angle=-90.]{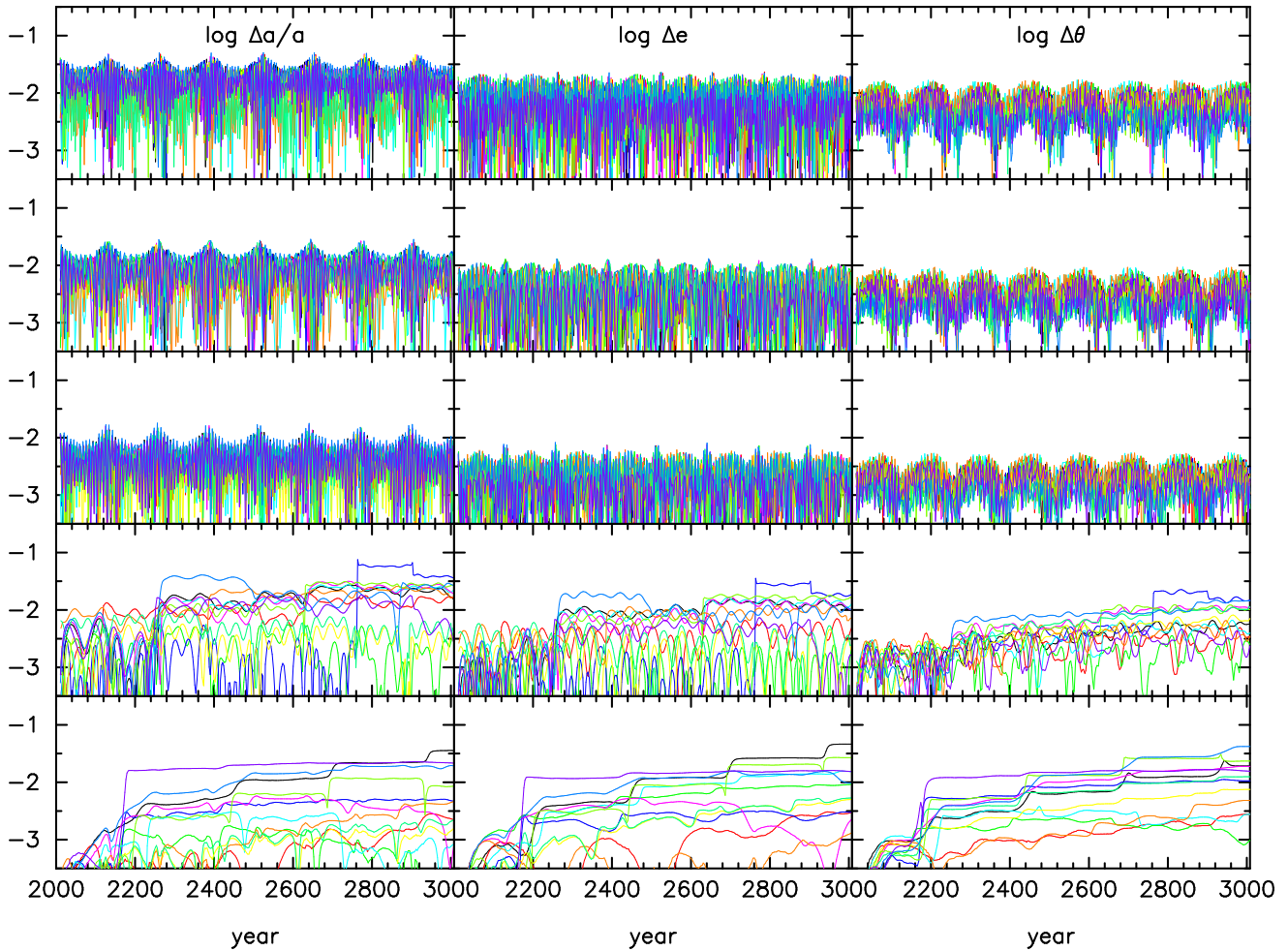}
  \end{center}
  \caption{Changes over the initial 1000 years (i.e. from 2008 to
    3008) of the orbital elements of star S1, in runs with
    $q=10^{-3}$, a circular IMBH orbit, and five different values of
    $a_{\rm IMBH}$, from $0.3\mpc$ (top) to $30\mpc$ (bottom).
    Different colors denote the 12 different orientations of the
    IMBH/SMBH orbital plane.}
  \label{fig:e0}
\end{figure*}

\begin{figure*}
  \begin{center}
    \includegraphics[width=13.cm,angle=-90.]{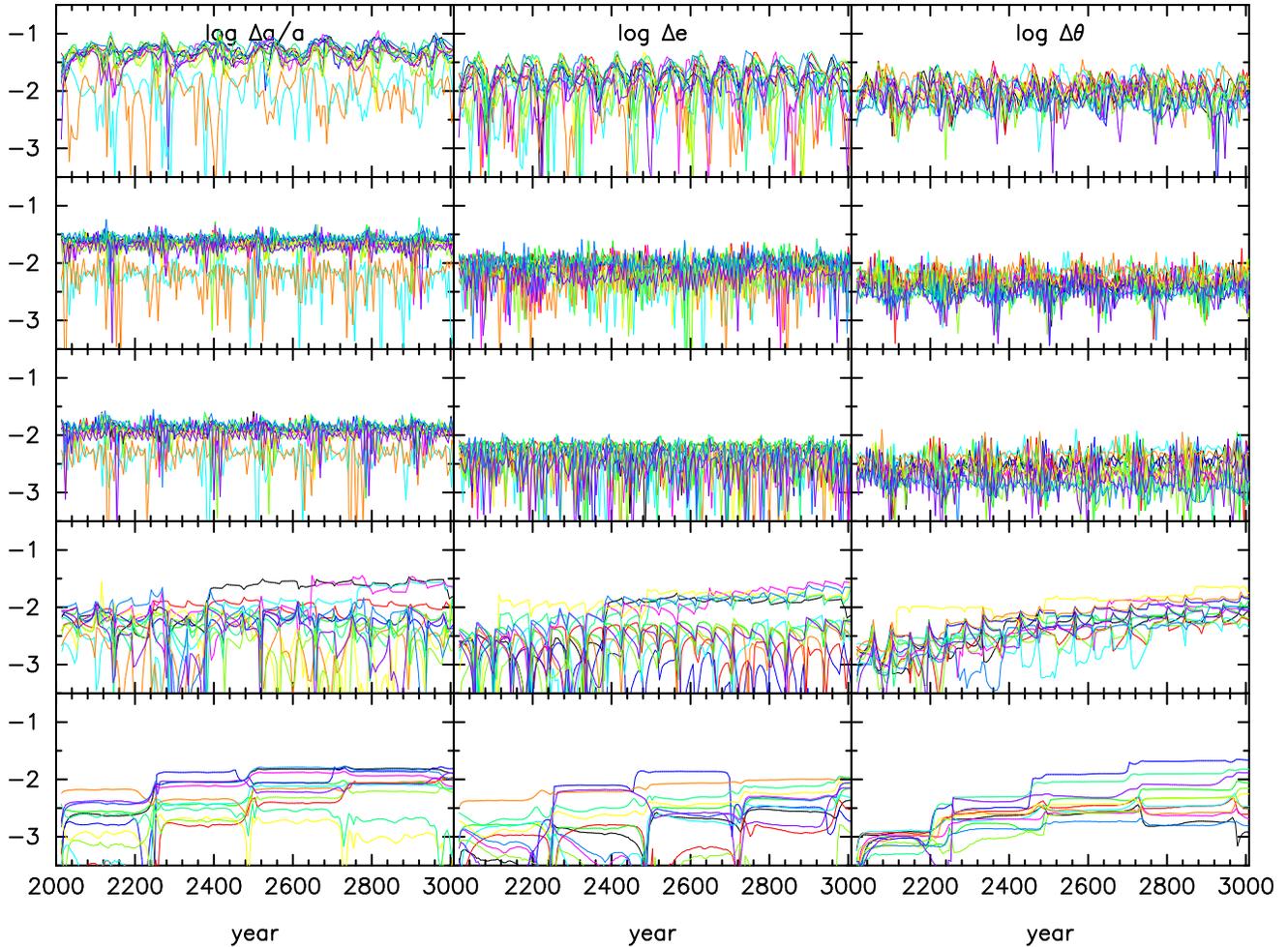}
  \end{center}
  \caption{Like Fig.~\ref{fig:e0}, but for runs with the
           eccentric IMBH/SMBH binary.}
  \label{fig:e0.7}
\end{figure*}

One motivation for basing our $N$-body models on the actual S-star
orbits is that we are able to make quantitative statements about how
these orbits would evolve in the near future.

While a close encounter of the IMBH with a star can produce a sudden
large change in the star's velocity, such encounters are relatively
rare, and the evolution of the star's trajectory is better described
in terms of the change with time of its orbital elements.
Figures~\ref{fig:e0} and~\ref{fig:e0.7} summarize the evolution of the
orbit of star S1 over the next $1000\yr$ in each of the 120
integrations corresponding to $q=10^{-3}$.  We have plotted the
changes in $a$, $e$ and $\theta$ with respect to their values in the
year 2008; here $\theta$ is the angle in radians between the star's
orbital angular momentum vector at the two times.

The IMBH lies completely inside the orbit of S1 in the
top three rows of Figure~\ref{fig:e0} and~\ref{fig:e0.7}.
In these runs,
the star's orbital elements (which are computed with respect to the
instantaneous position and velocity of the SMBH) are found to
oscillate due the movement of the SMBH in response to the IMBH.
Oscillation on both the (short) timescale of the IMBH/SMBH orbit, and
the (longer) timescale of the star's orbit, can be seen.
The amplitude of these oscillations is greatly reduced when
the orbital elements are computed with respect to the center of mass of the 
IMBH/SMBH binary; there is almost no true evolution of the star's orbit on
these time scales.

When the IMBH's orbit is larger than that of the star, oscillatory
changes in the star's orbit still take place on the (now shorter) timescale
of its own orbit, but in addition there are stepwise changes that
occur roughly once per orbit of the IMBH, when it comes closest to the
star.  The latter changes are similar in magnitude for the circular
and eccentric binaries, but in the latter case, changes in S1's orbit
are less dependent on the binary's orientation.

We chose to place the IMBH at apoastron at the start of the integrations,
and this choice determined the time of first and subsequent ``close
encounters'' with each star.  This is clear in the lower panels of
Figs.~\ref{fig:e0} and~\ref{fig:e0.7}, where the IMBH orbit has a
longer period than that of S1.  Had we varied the argument of
periastron for the IMBH, the jumps in S1's orbital elements would have
taken place at different times with respect to 2008.  
These jumps can be significant,
e.g. as large as $\sim 30$ arc minutes in the case of $\theta$.
However based on the figures, the a priori chance of such a jump
occurring during any short ($\sim$ years-long) interval of time is
small.

According to \citet{gill09}, the errors in the observationally
determined values of $a$, $e$ and $\theta$ for S1 are $\sim 5\%$,
$\sim 0.03$ and $\sim 0.01$ radians respectively.  Figures~\ref{fig:e0} 
and~\ref{fig:e0.7} suggest that short-term ($\sim$ few years) changes
in the semi-major axis might rise above the measurement
uncertainties for IMBH/SMBH binaries with $q\approx 10^{-3}$ and
$a\lesssim 1\mpc$ ($e=0$) and $a\lesssim 3\mpc$ ($e=0.7$).  Changes in
$e$ and $\theta$ might barely be detectable on the same timescale for
$a\lesssim 0.3\mpc$.

\section{Results: Long-term evolution}
The perturbations exerted by the IMBH induce changes in the orbital
elements of the stars. In particular, the presence of an IMBH has the
following effects: (i) randomization of the inclinations of the stars, (ii)
ejection of stars from the region, (iii) scattering of stars onto plunging 
orbits that result in tidal disruption in the SMBH's tidal field (iv) 
secular effects like Kozai cycles.

\subsection{Randomization of the orbital planes}
\begin{figure}
  \begin{center}
    \includegraphics[width=6.5cm, angle=270]{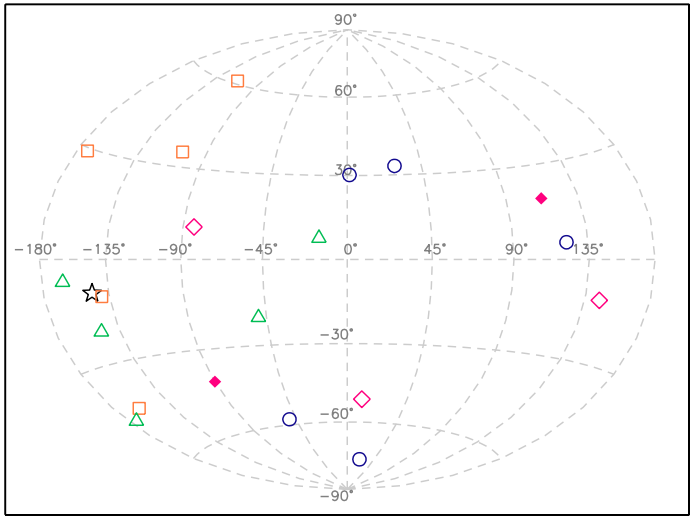}
  \end{center}
  \caption{Orientation of the orbital plane of star S9 as currently
    observed (star) and at the end of the simulations (empty symbols
    if bound star; full symbols if unbound star). Different symbols
    are for different binary mass ratios: $q=1.0\times10^{-4}$
    (circles), $q=2.5\times10^{-4}$ (squares), $q=5.0\times10^{-4}$
    (triangles), $q=1.0\times10^{-3}$ (diamonds).}
  \label{fig:map8}
\end{figure}

The randomization of the orbital planes has been shown by
\citet{mgm09} to be efficient for IMBH masses larger than $\sim
1500\msun$ and eccentricities larger than $\sim 0.5$. In these cases,
a thin corotating disk of stars was converted into an isotropic
distribution in just $\sim 1\myr$. In this work, we consider a
population of stars that are already approximately isotropic at the
beginning of the simulations. Nonetheless, the IMBH acts to randomize
the orbital planes of the stars so that the angular momentum vectors
can point to very different directions at the end of the $10\myr$
evolution, as can be seen in Figure\,\ref{fig:map8} in the case of star
S9.

\subsection{Ejections}
Perturbations from the IMBH can also result in ejections of stars from
the region. While ejections can be produced by close encounters with
the IMBH, the majority of the events are due to slow but steady
increase in a star's eccentricity, until eventually the star becomes
unbound to the SMBH. We expect the probability of ejection to depend
on both the stellar and the IMBH orbital parameters.
\begin{figure*}
  \begin{center}
    \includegraphics[width=8.0cm]{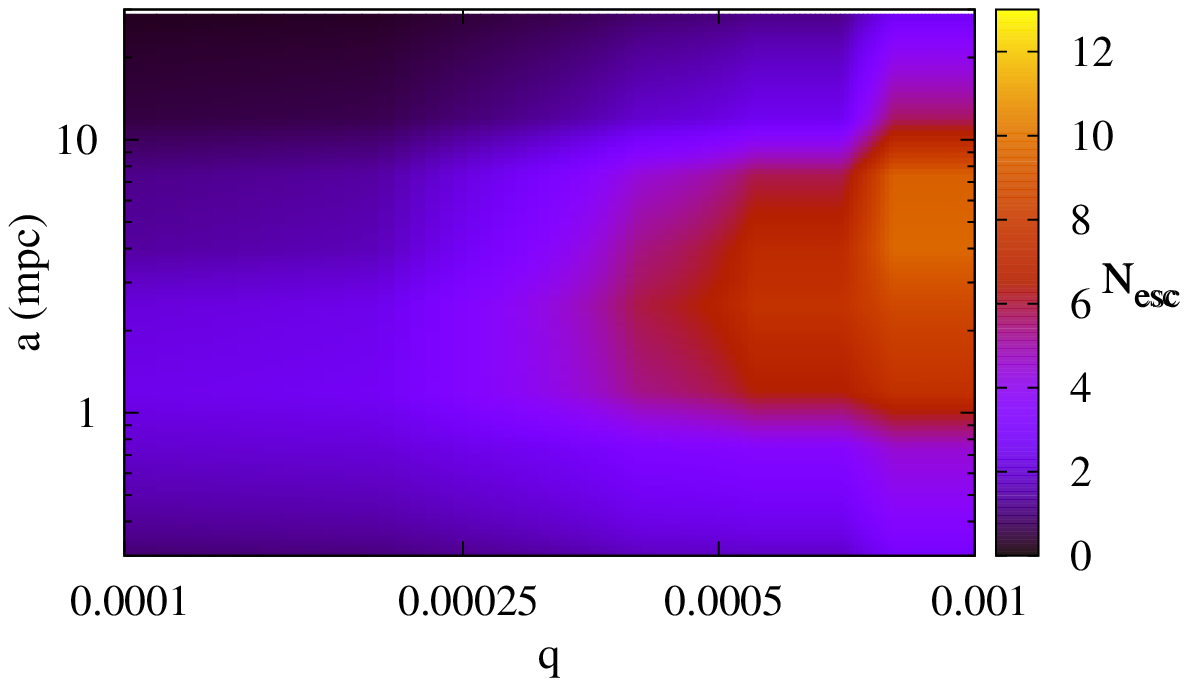}
    \includegraphics[width=8.0cm]{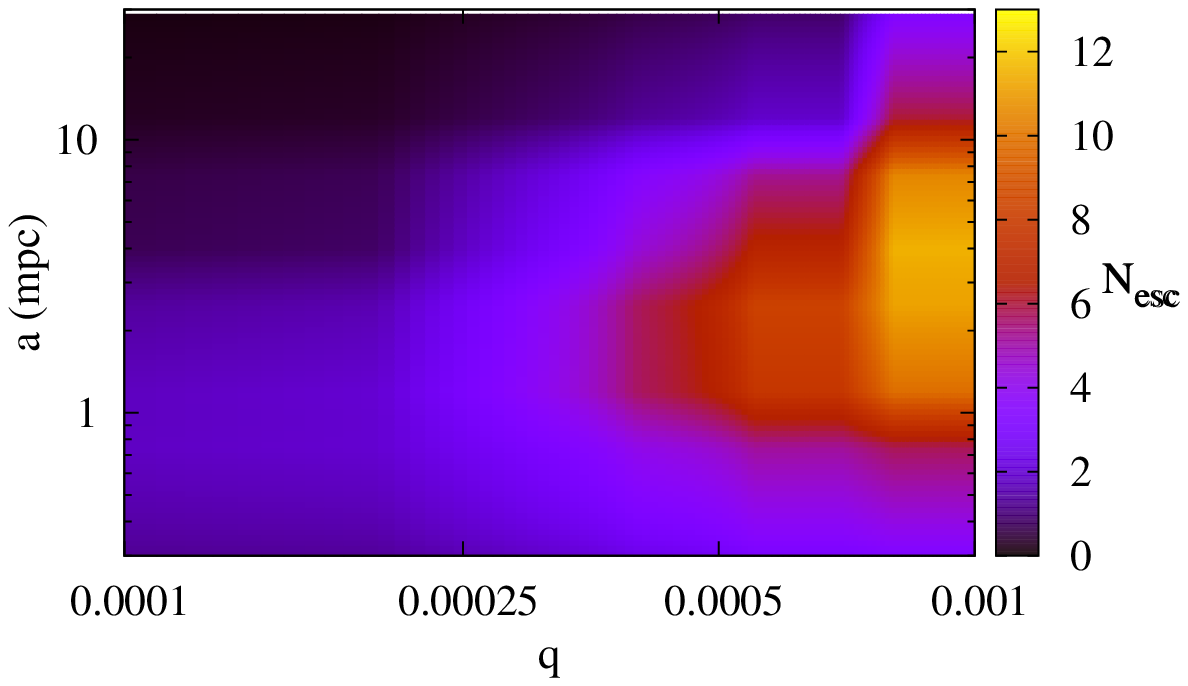}
  \end{center}
  \caption{Average number of ejected stars as a function
    of the black hole binary parameters, in the case of an initially circular
    (left) and eccentric (right) binary.}
  \label{fig:enesc}
\end{figure*}
Figure\,\ref{fig:enesc} shows the average number of escapers produced in
any simulation with given mass ratio $q = \mim/\mbh$ and semi-major
axis $a$ of the binary black hole. We find that most of the ejections
occur for large mass ratios and for separations comparable to the
typical distances of the S-stars. These parameters maximize the
strength of the perturbations, which tend to alter the eccentricity of
the star.  A circular and an eccentric binary seem to produce
approximately the same number of escapers.

\begin{figure}
  \begin{center}
    \includegraphics[width=6.0cm,angle=270]{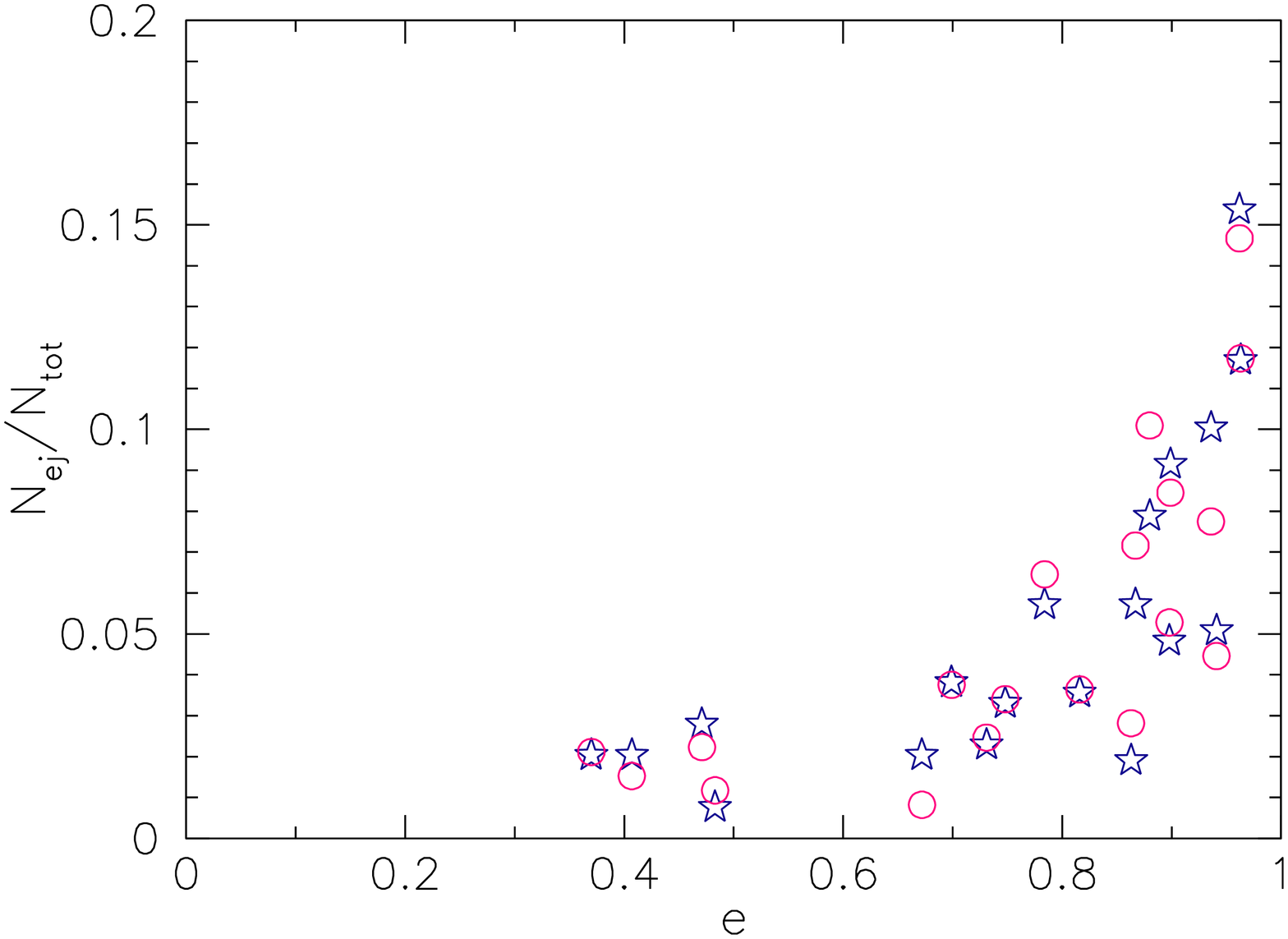}
  \end{center}
  \caption{Fraction of ejections experienced by each S-star as a
    function of the star's initial eccentricity in the case of a
    circular (stars) and eccentric (circles) binary.}
  \label{fig:nef}
\end{figure}

Stars with initially large eccentricity are more susceptible to
perturbations and therefore more likely to be ejected, as shown in
Figure\,\ref{fig:nef}. In particular, with $e = 0.96$ star S14 is the
one which experiences the most ejections.

Extrapolating the results obtained for the S-stars sample to the real
Galactic center population (assuming a total of $10^4$ stars in the
same region) we obtain an ejection rate $\mathcal{R} \sim
1.8\times10^{-4}\yr^{-1}$.

\subsubsection{Survival time}
\begin{figure*}
  \begin{center}
    \includegraphics[width=4.0cm]{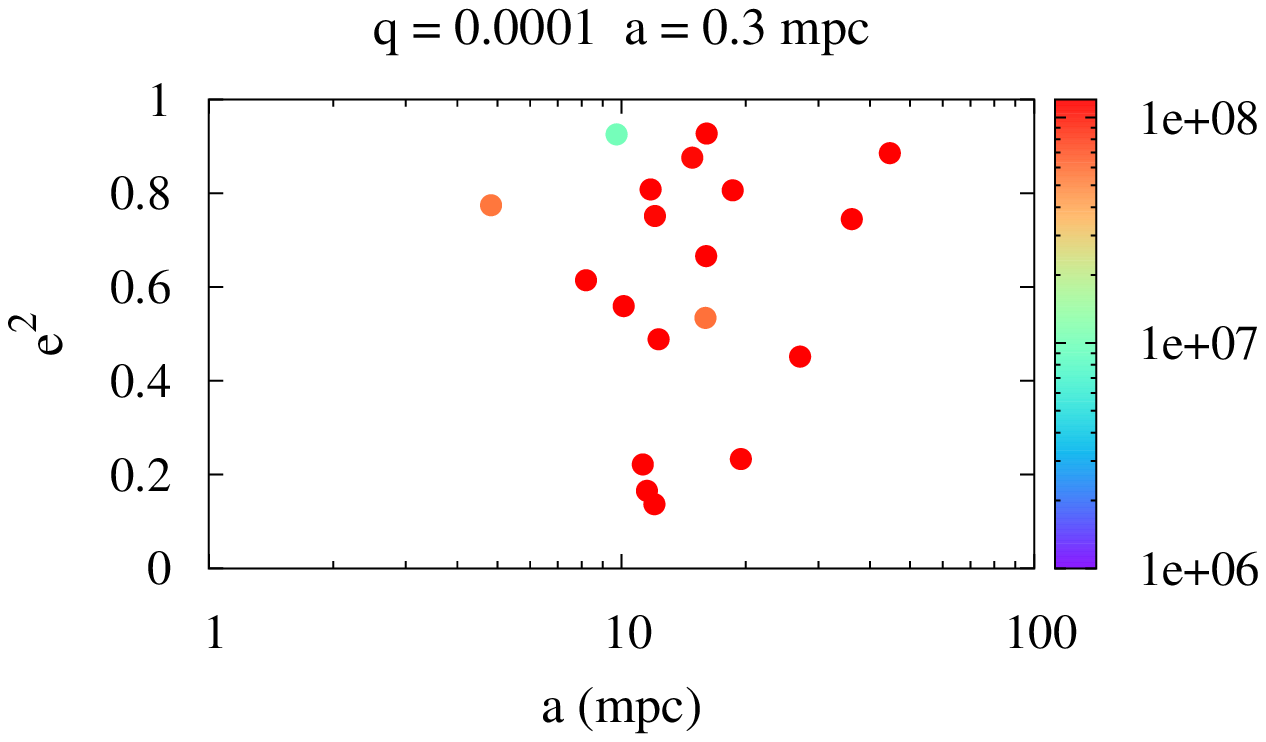}
    \includegraphics[width=4.0cm]{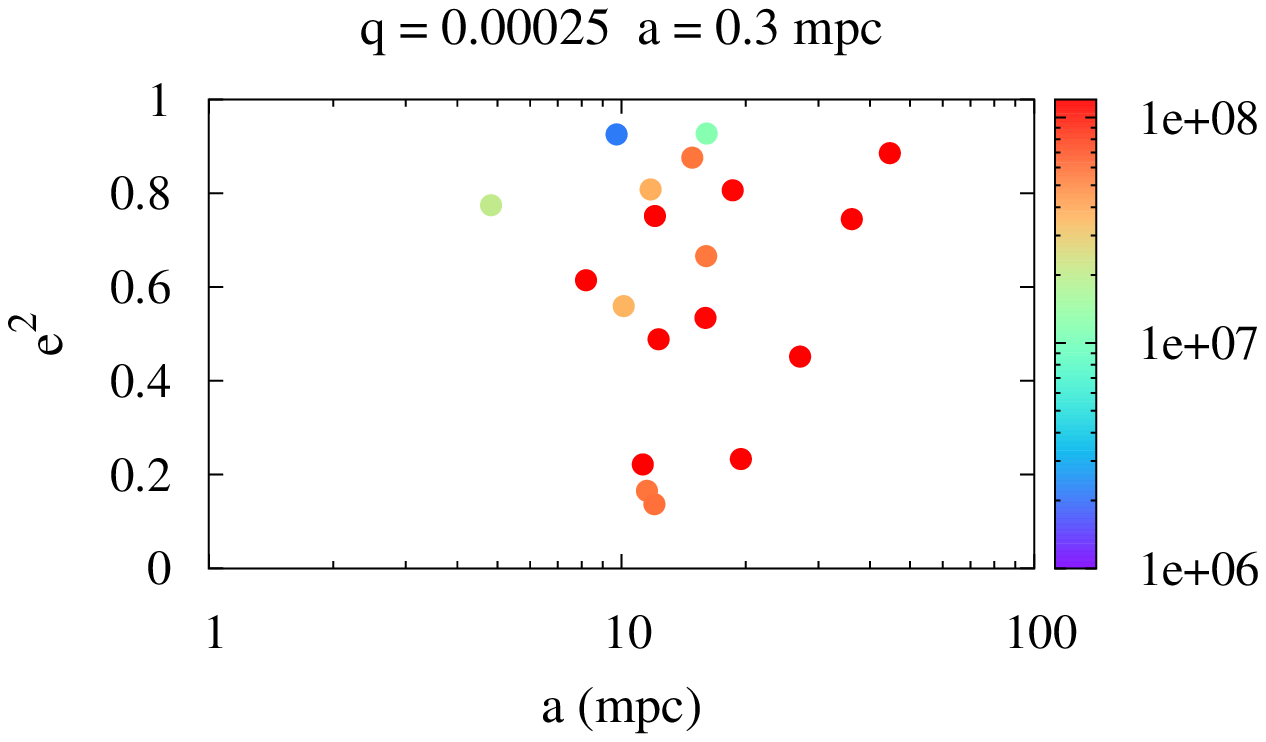}
    \includegraphics[width=4.0cm]{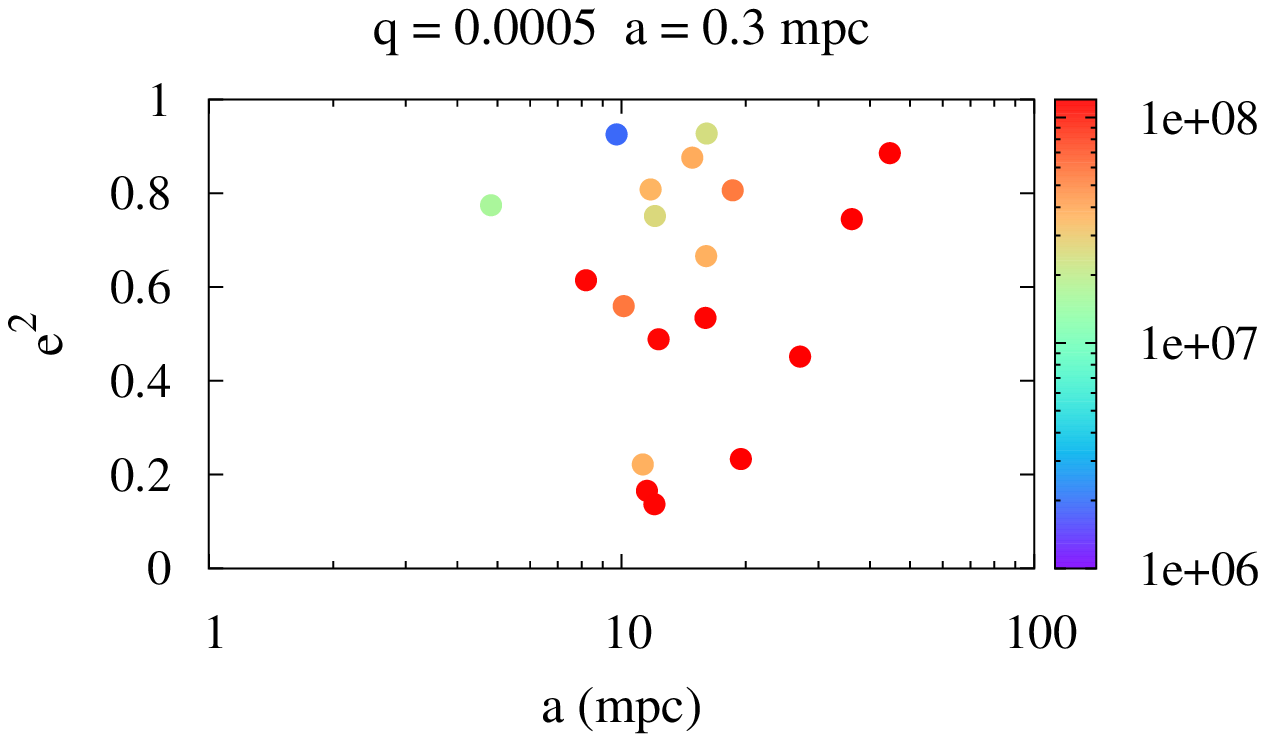}
    \includegraphics[width=4.0cm]{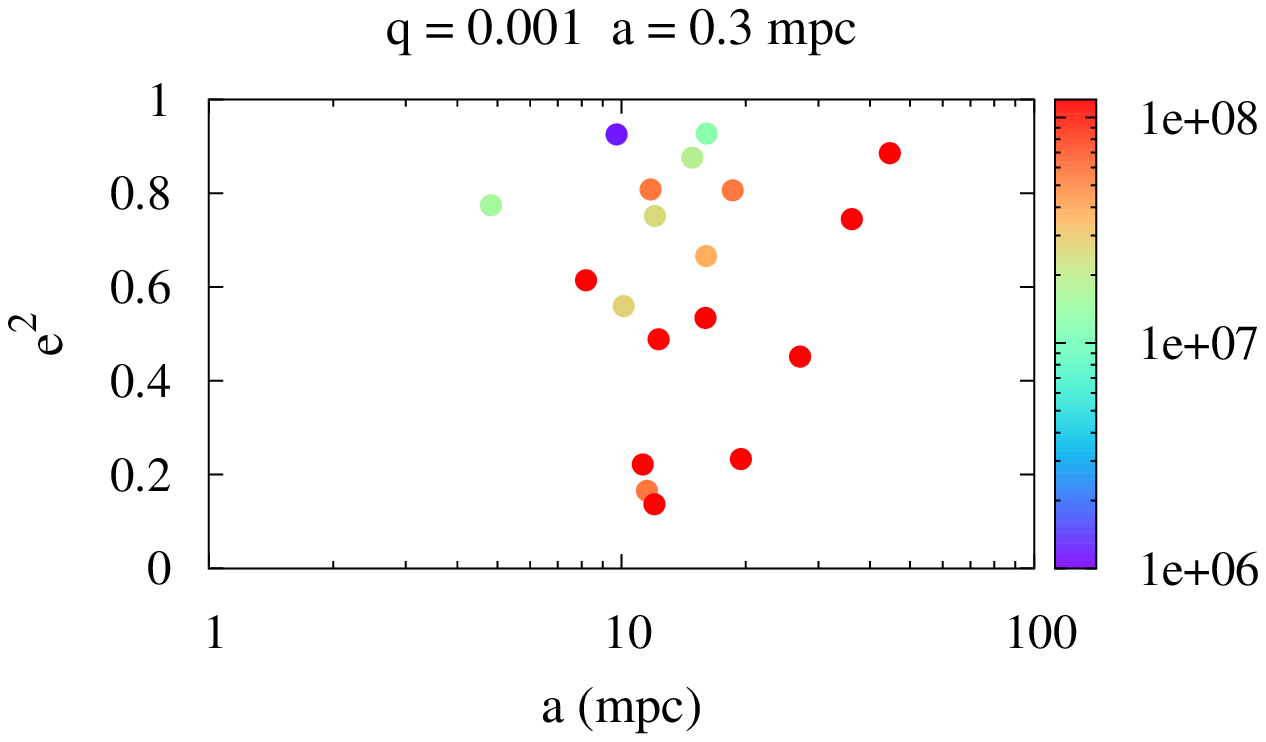}
    \includegraphics[width=4.0cm]{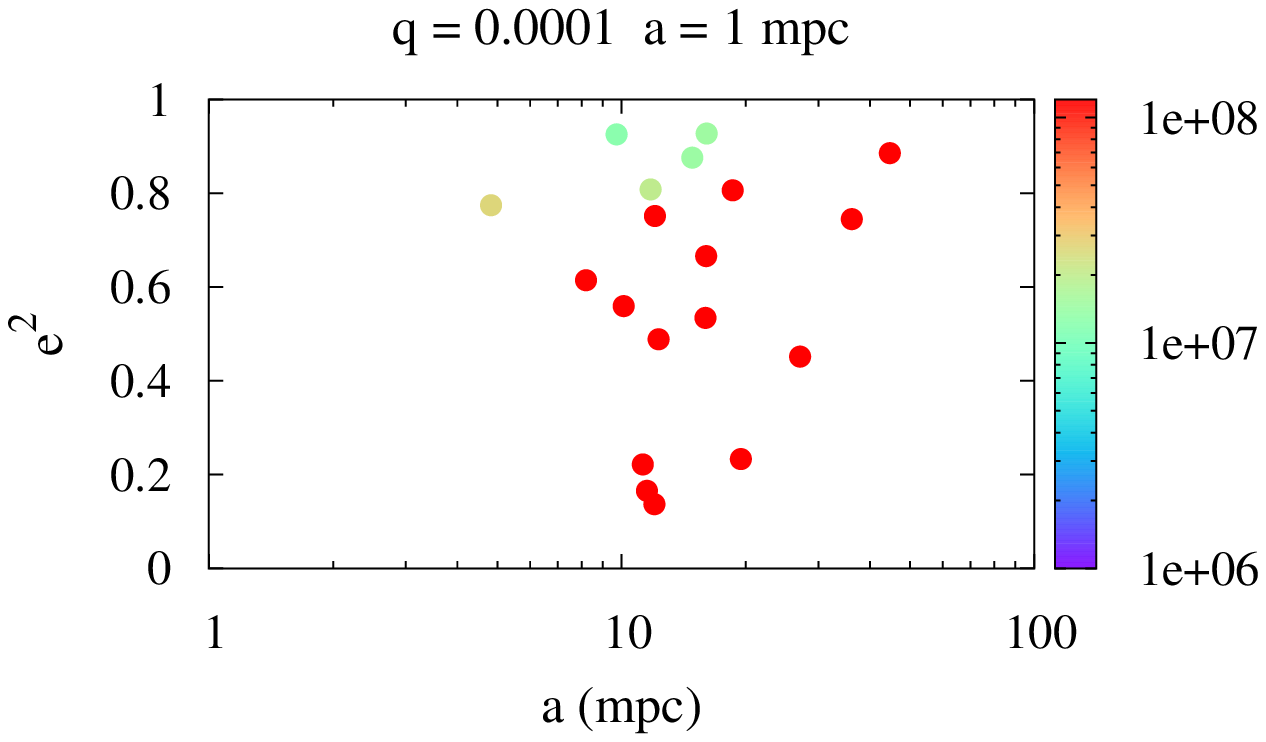}
    \includegraphics[width=4.0cm]{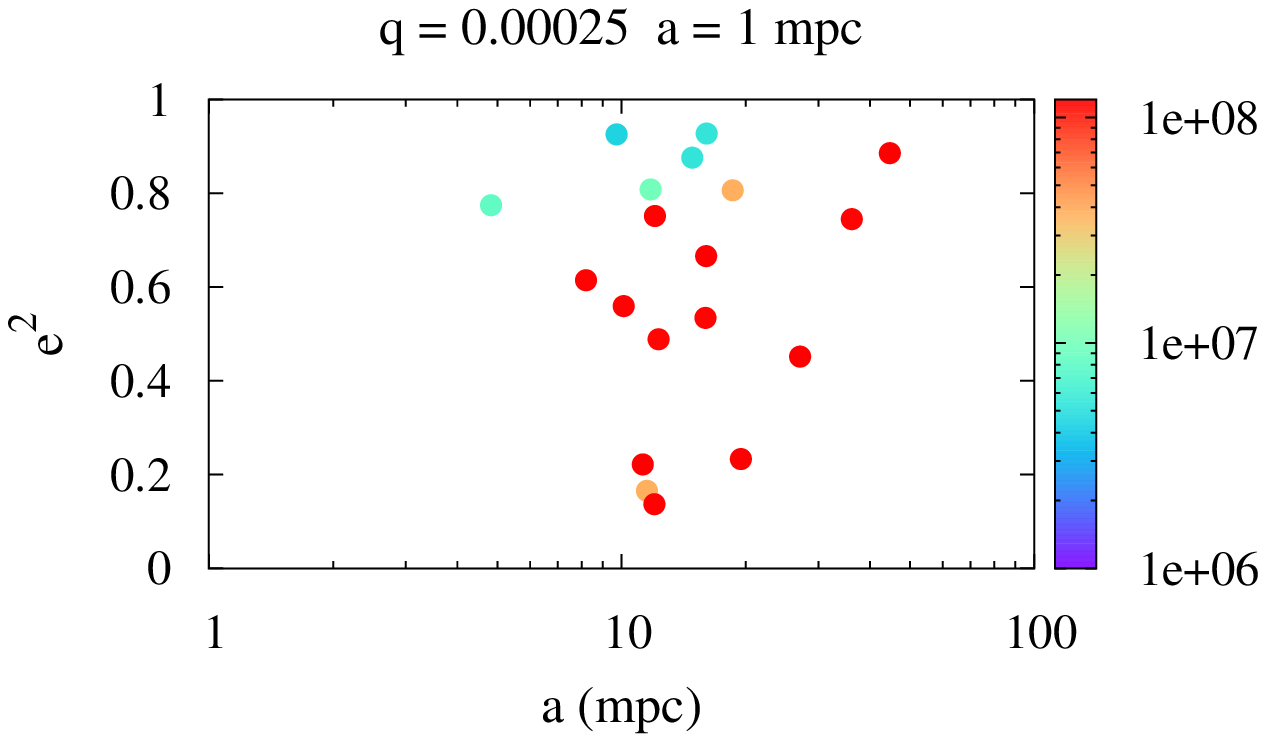}
    \includegraphics[width=4.0cm]{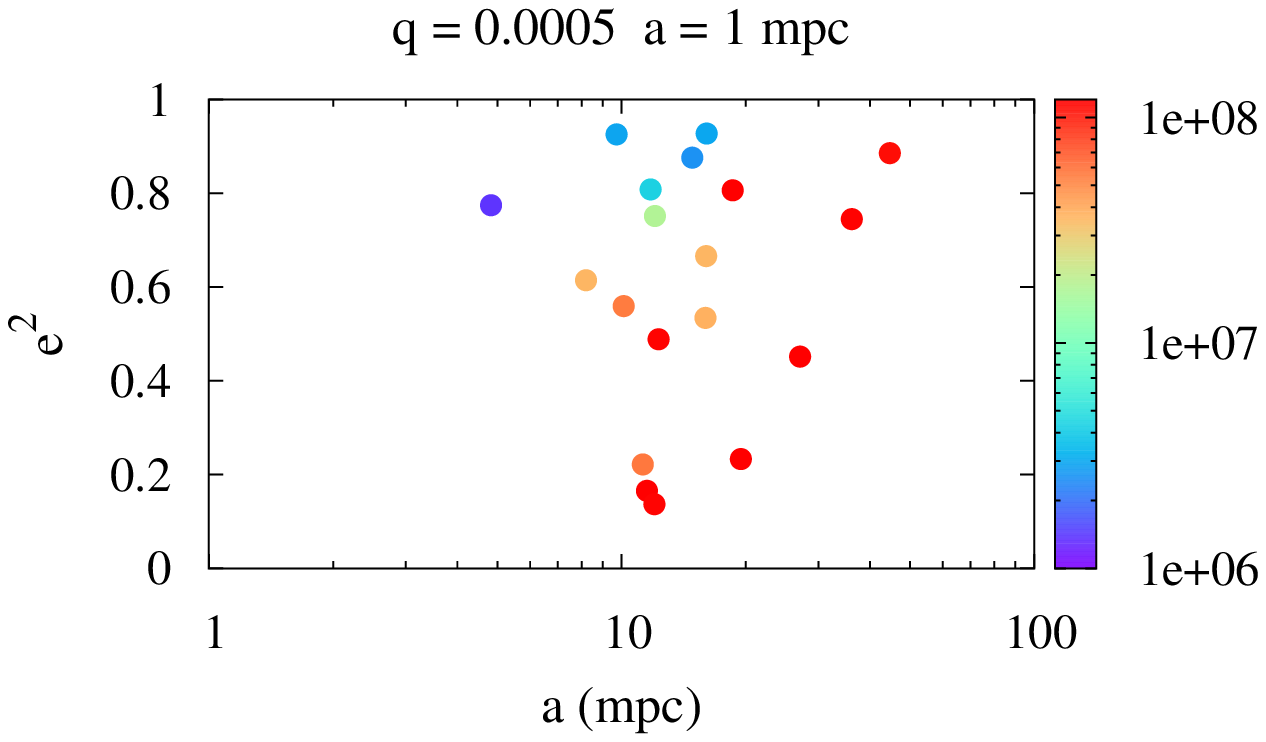}
    \includegraphics[width=4.0cm]{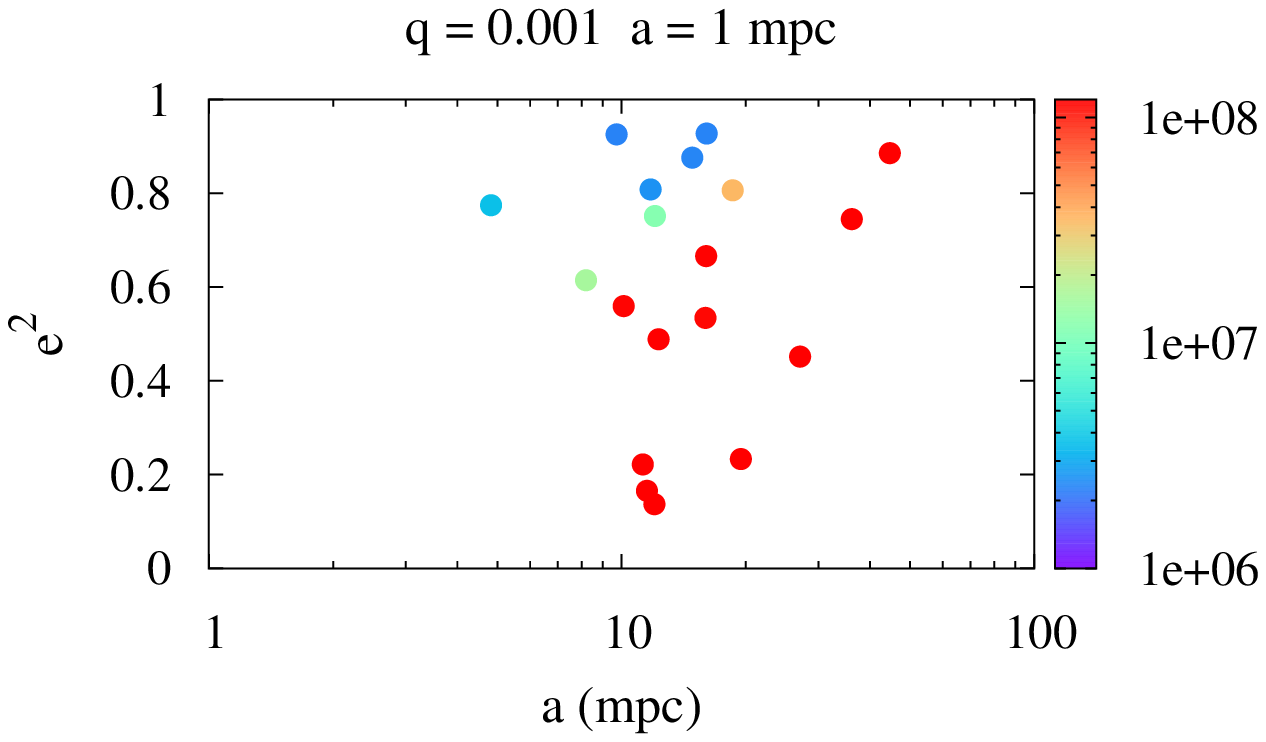}
    \includegraphics[width=4.0cm]{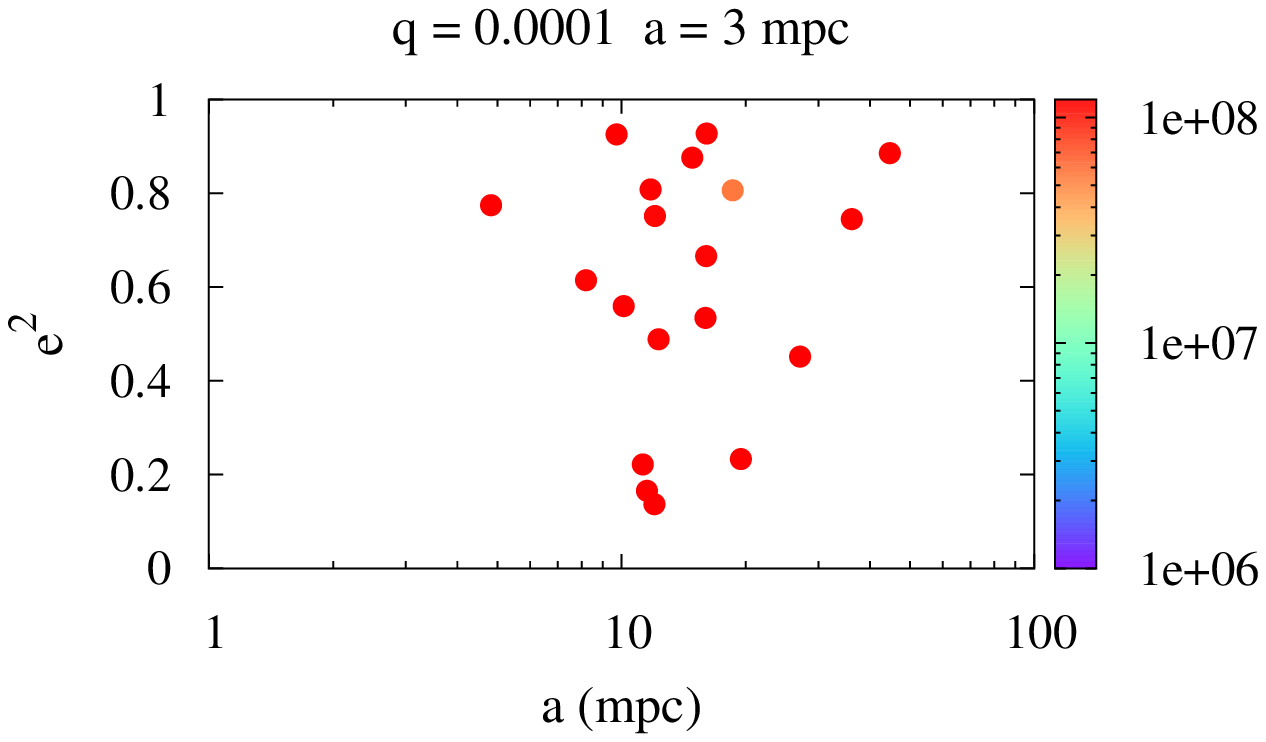}
    \includegraphics[width=4.0cm]{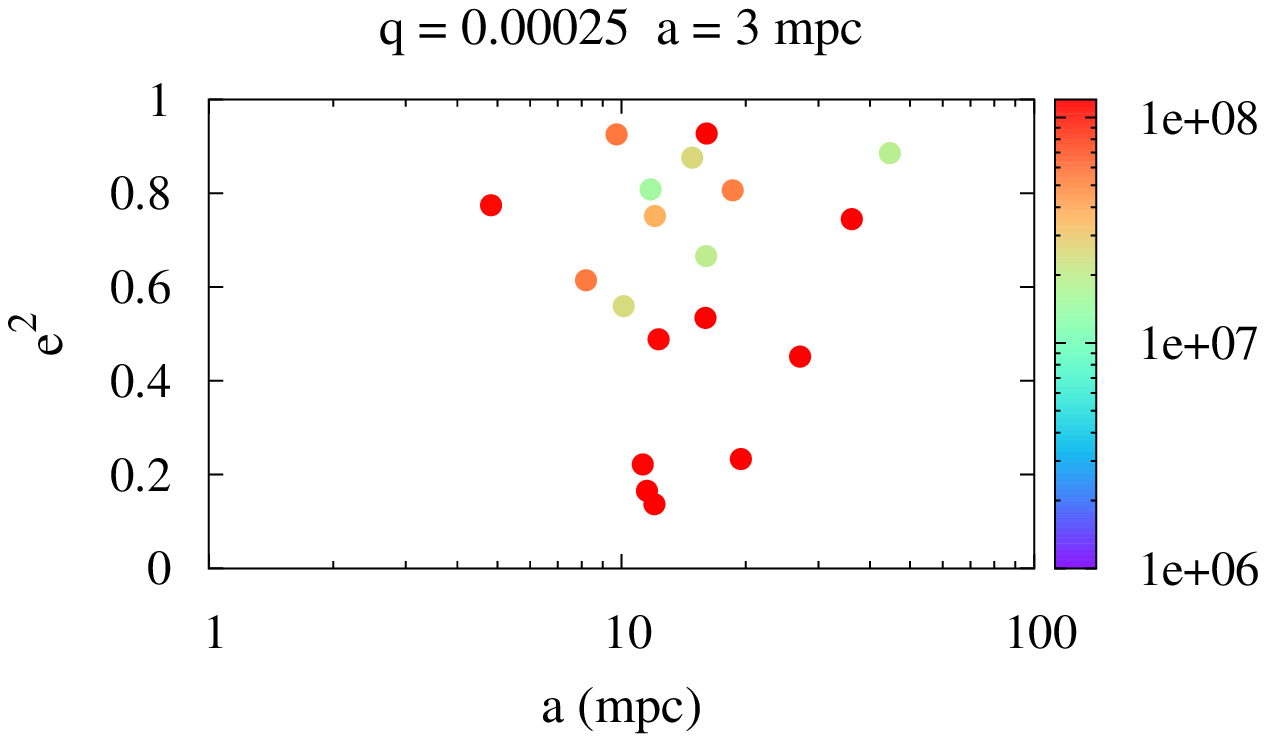}
    \includegraphics[width=4.0cm]{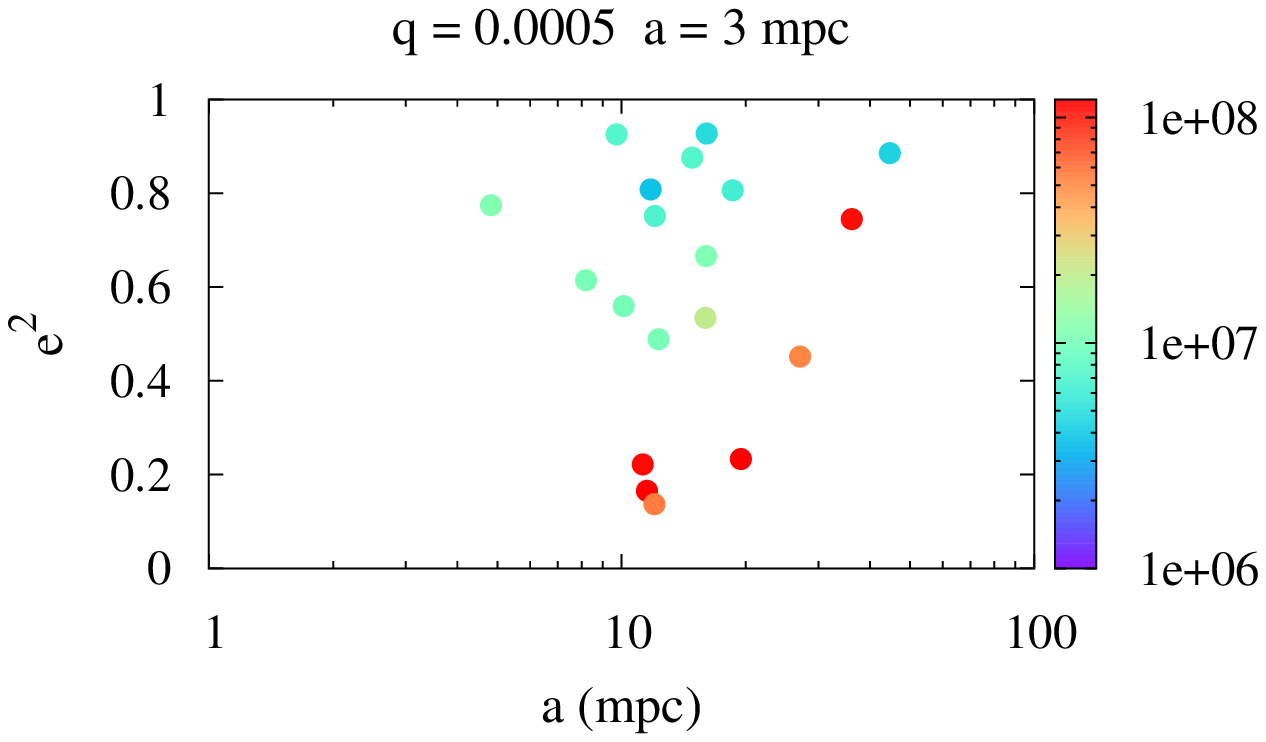}
    \includegraphics[width=4.0cm]{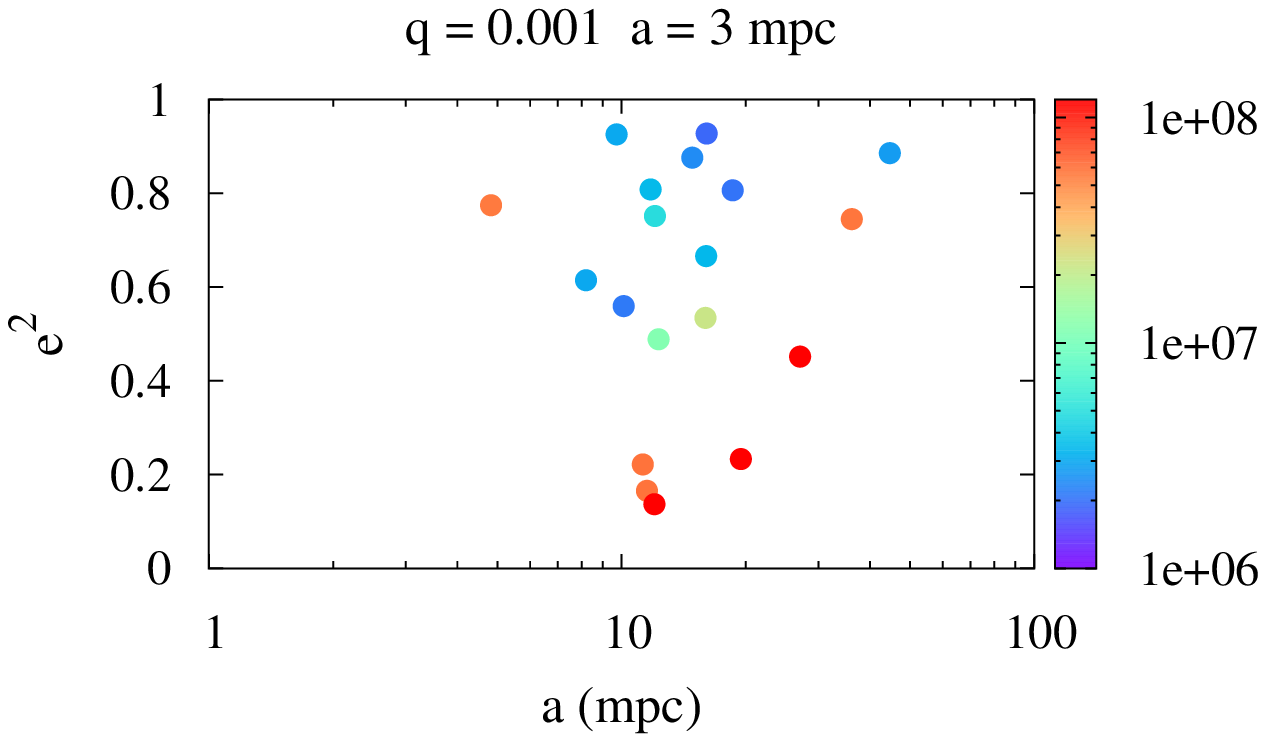}
    \includegraphics[width=4.0cm]{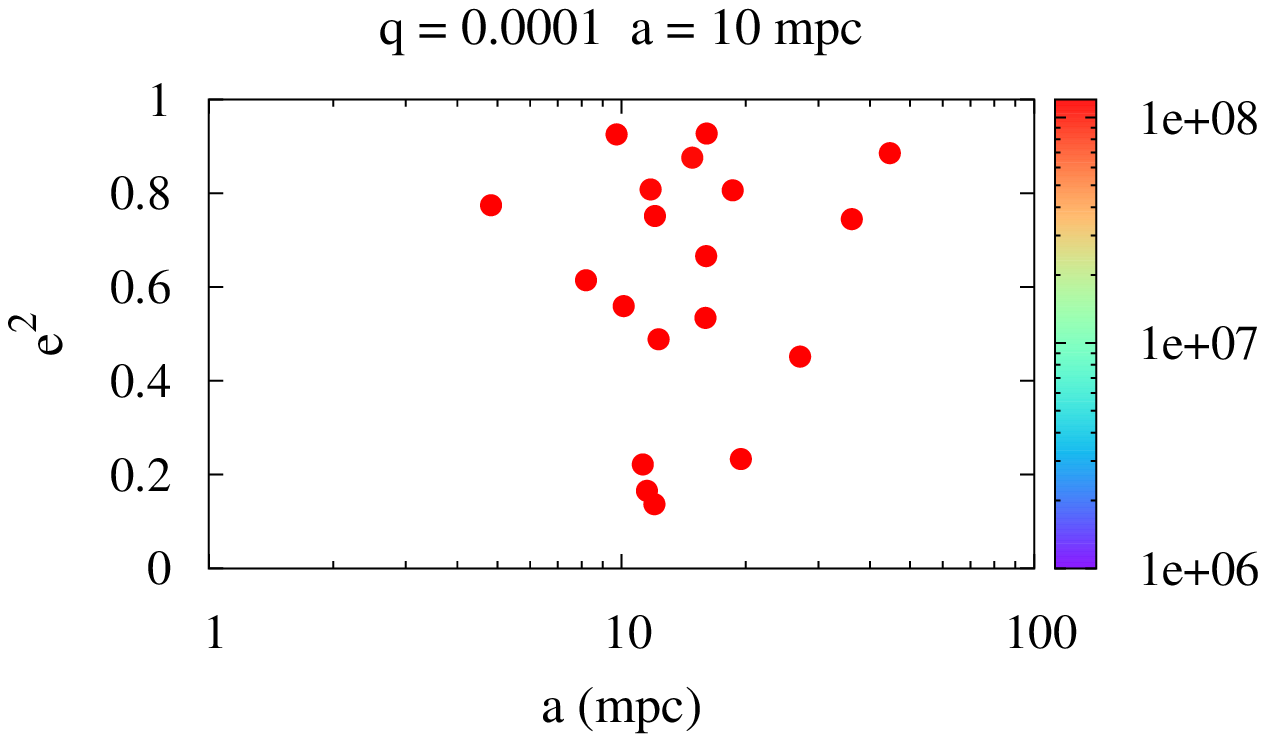}
    \includegraphics[width=4.0cm]{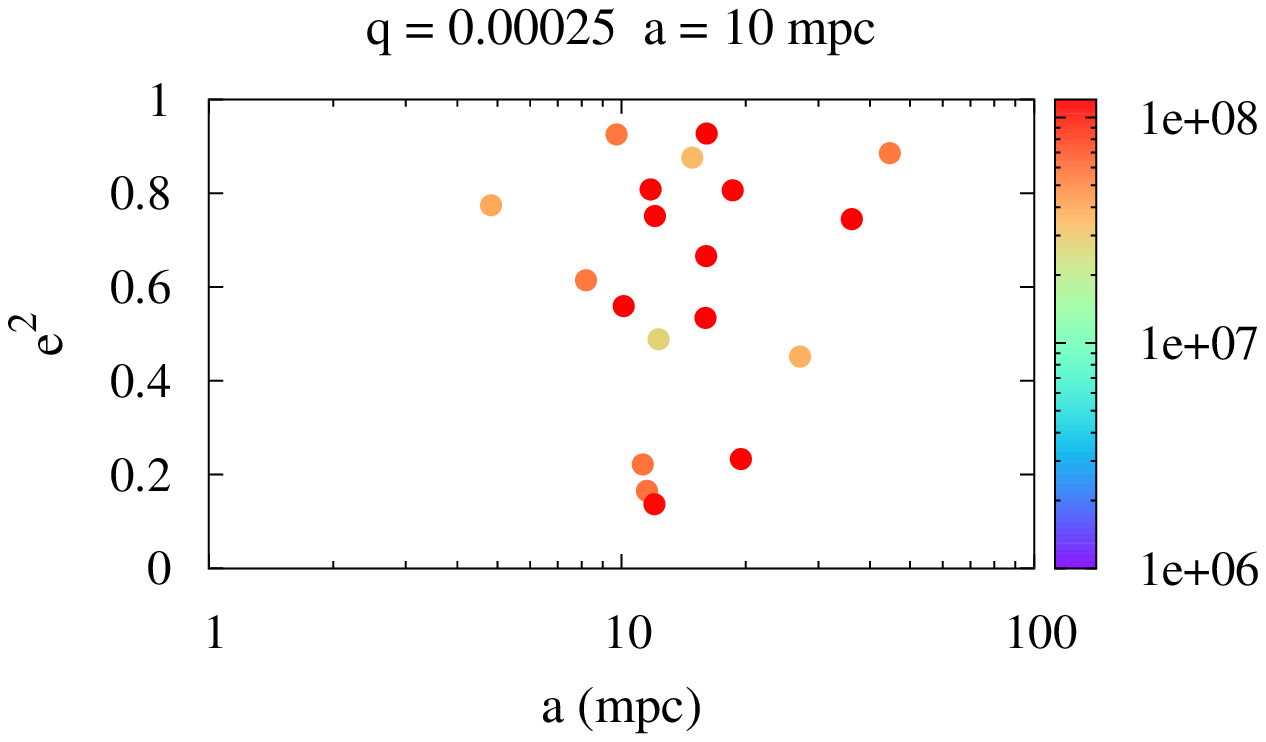}
    \includegraphics[width=4.0cm]{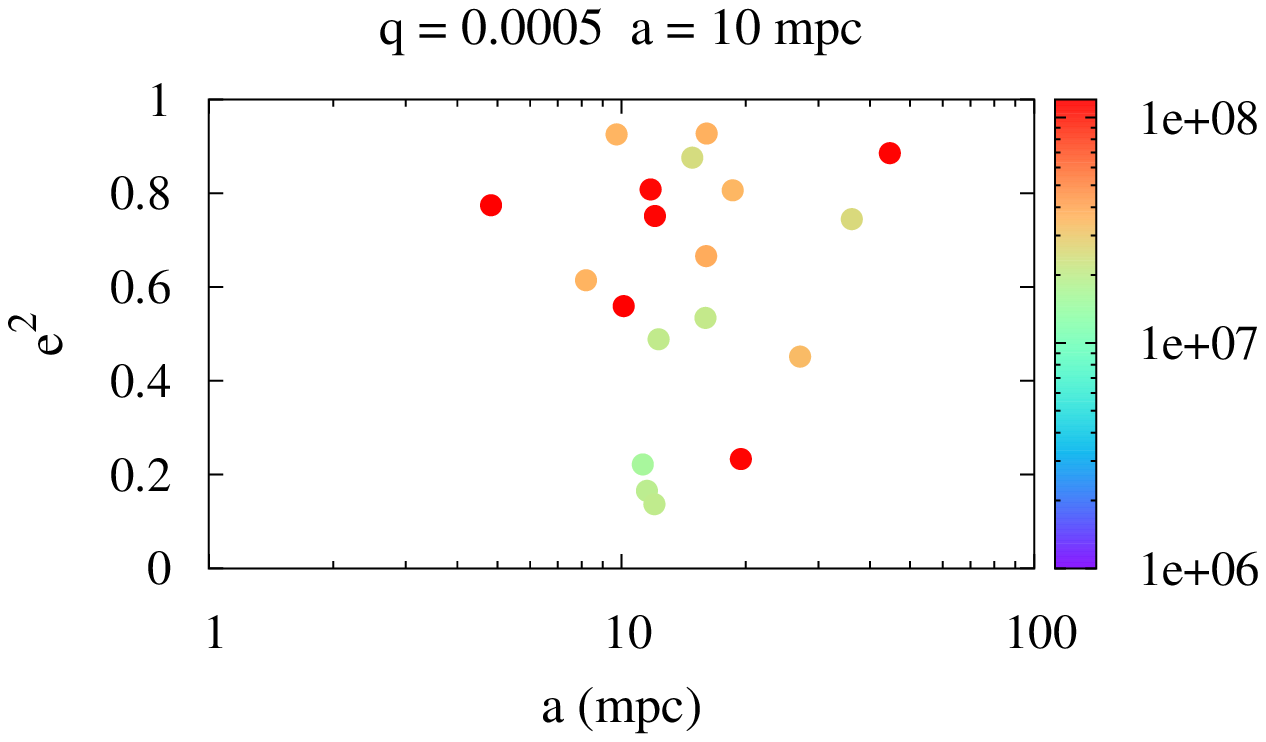}
    \includegraphics[width=4.0cm]{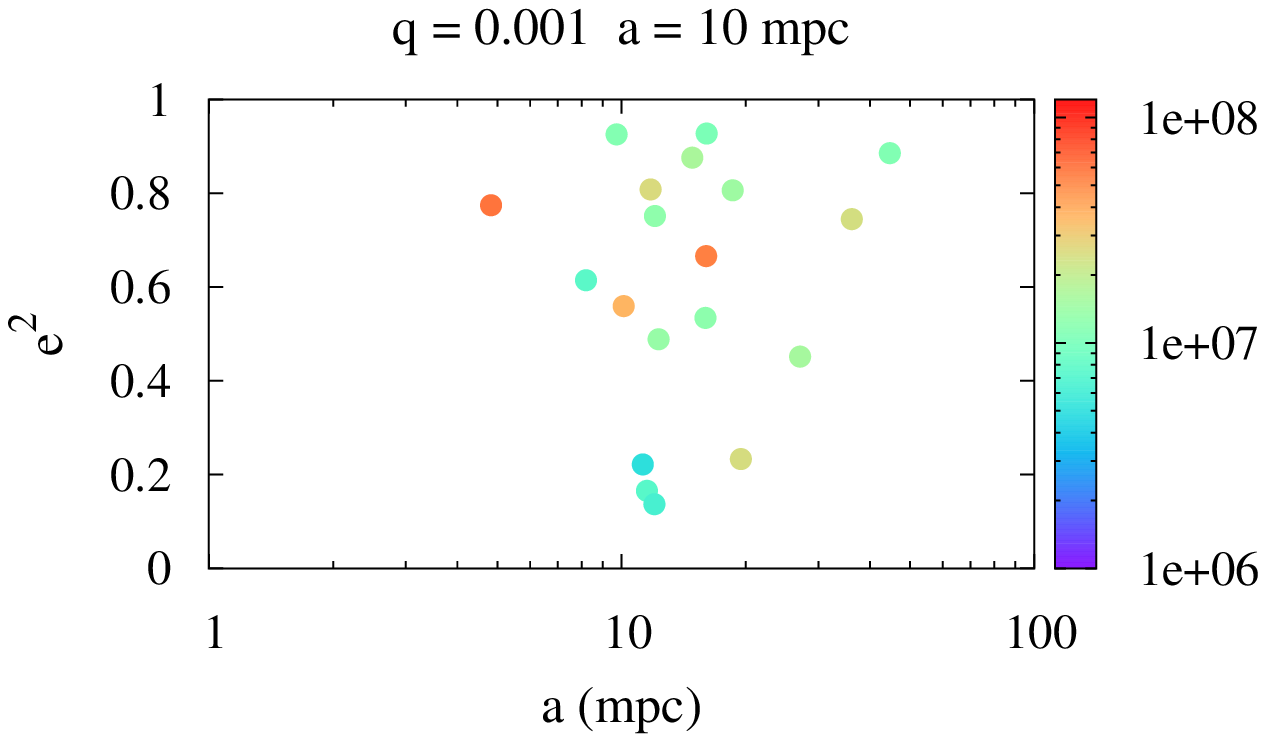}
    \includegraphics[width=4.0cm]{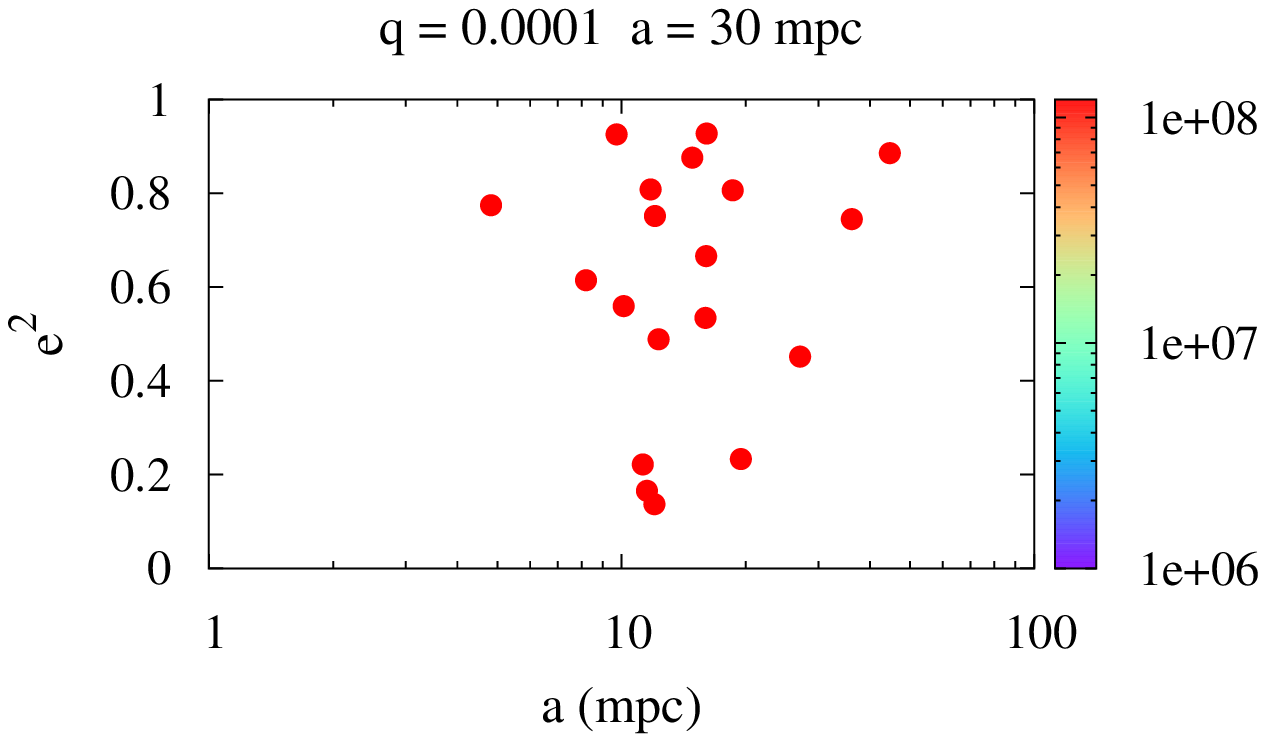}
    \includegraphics[width=4.0cm]{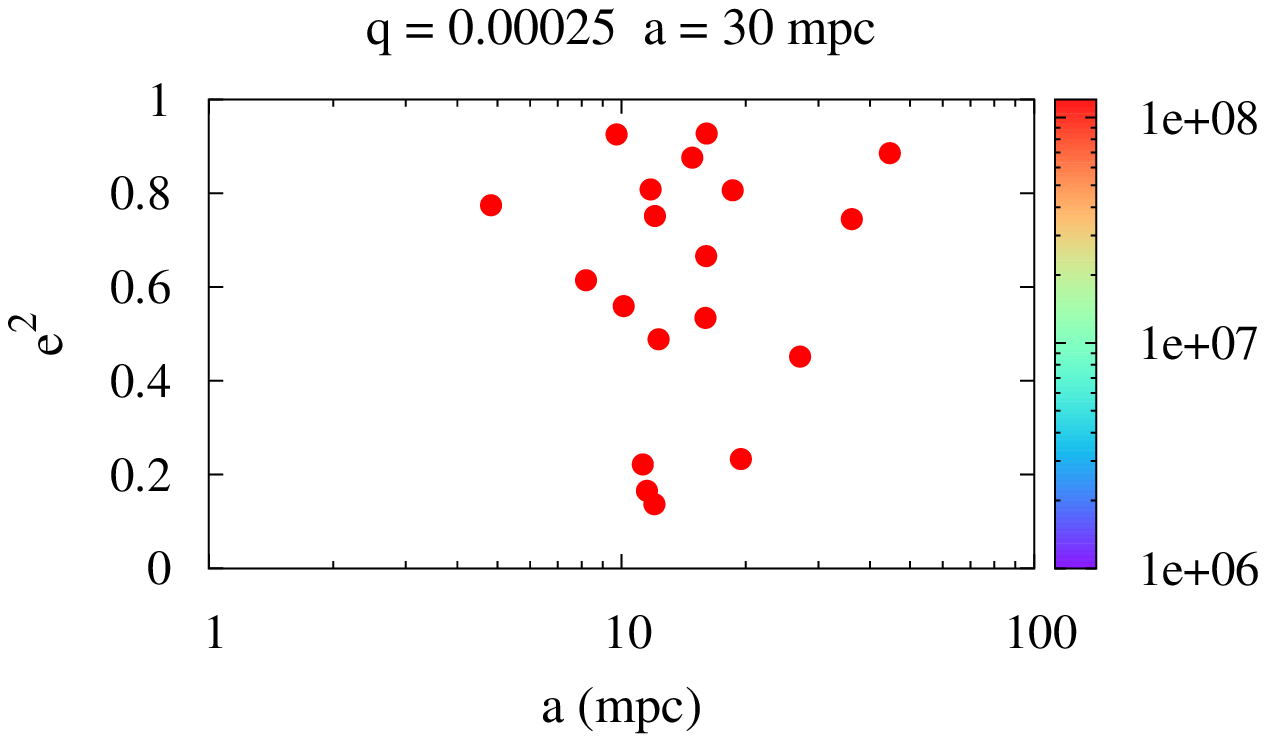}
    \includegraphics[width=4.0cm]{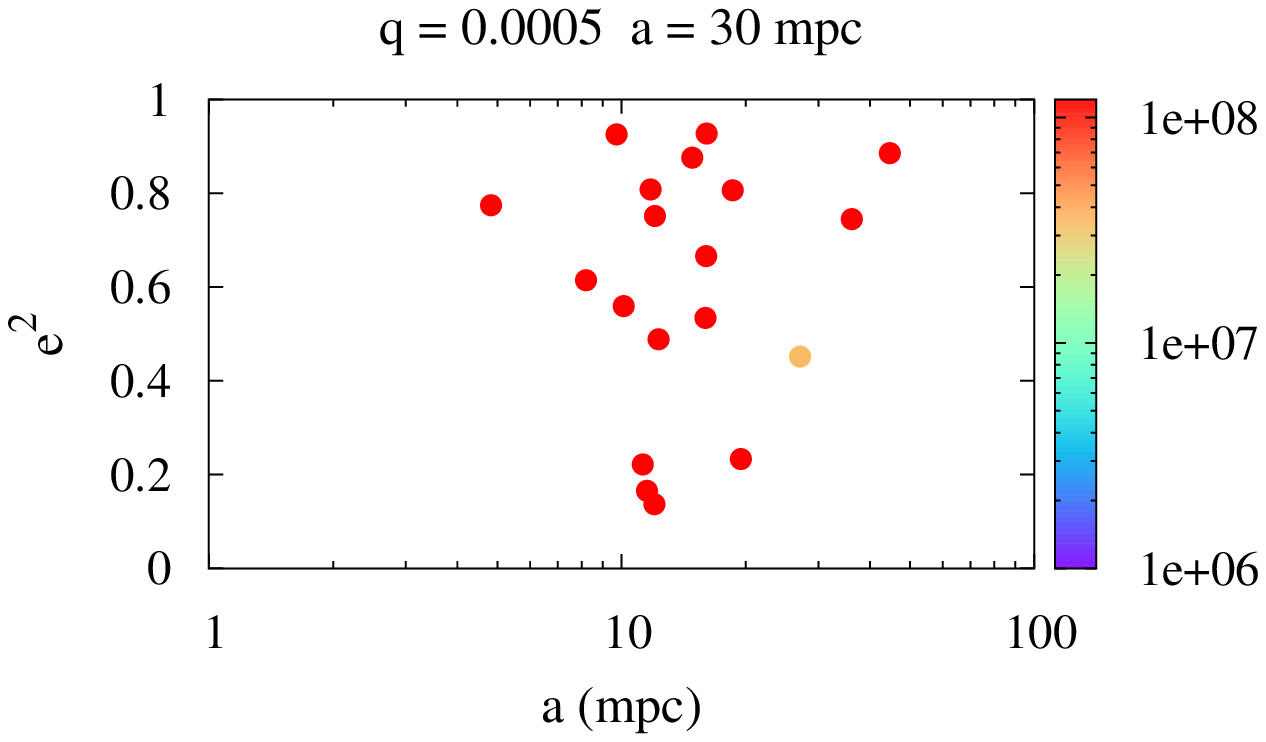}
    \includegraphics[width=4.0cm]{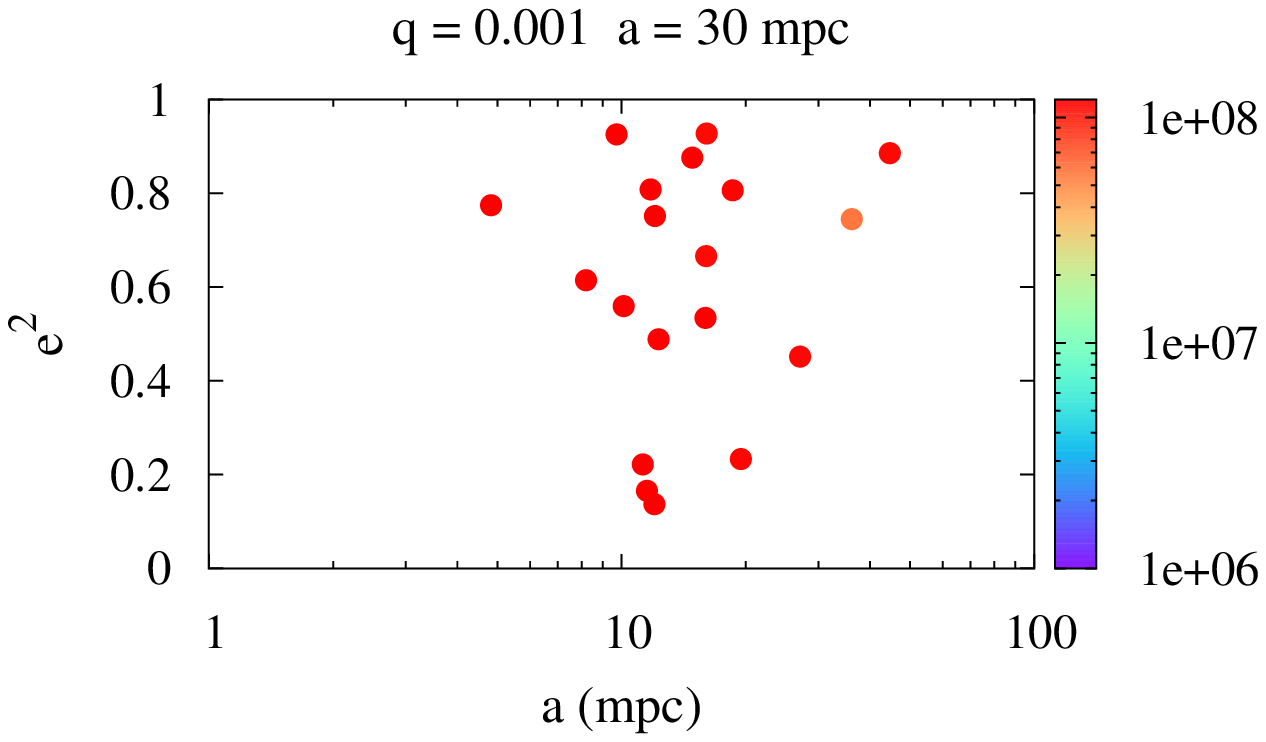}
  \end{center}
  \caption{Mean time until ejection for all the stars as a function of
    the binary orbital parameters.}
  \label{fig:tmean}
\end{figure*}

Since stars can become unbound from the SMBH and escape the region
monitored by observations, it is interesting to study the ``survival''
probability for the S-stars, for a given set of orbital parameters. We
estimate the mean time until ejection $T_{ej}$ for individual stars
for a given set of binary parameters. Since not all stars are ejected
before the end of the simulation, an average of the ejection time over
the cases where ejection occurs would yield a lower limit to
$T_{ej}$. The problem of estimating the lifetime of a population given
a discrete set of events within a finite time is called ``survival
analysis under censoring'', where censoring refers to the fact that at
the end of the simulation only a fraction of the stars have a measured
time of ejection. We therefore apply a maximum likelihood statistical
analysis to our sample of censored data and compute the mean ejection
time for each star and for a given set of binary parameters by
averaging over the 12 runs which correspond to the different binary
orientations in the sky.  We define $N_e$ the number of ejection
events experienced by a star by the end of the simulations at time $T$
and $N_c$ the number of non ejections.  If $t_i, i=0...N_e$ are the
ejection times for the events, the mean ejection time for a given star
is
\begin{equation}
T_{\rm ej} = \frac{1}{N_e} \sum_{i = 0}^{N_e} t_i  + \frac{N_c}{N_e} T
\end{equation} 
The results are shown in Figure\,\ref{fig:tmean} for all stars and
provide a quantitative estimate of the probability of ejection for
individual stars taking into account both the orbital properties of
the stars and of the IMBH.
The mean time to ejection is $\simless 10^7\yr$ in the case of binaries
with $q\simgreat5.0\times10^{-4}$ and $3\mpc \simless a \simless
10\mpc$. Similarly to Figure\,\ref{fig:nef}, stars with large
eccentricities have the shortest $T_{\rm ej}$.

\subsubsection{Hypervelocity stars}
A fraction of the escapers produced during the evolution can be
ejected with large enough velocity to become unbound to the Galaxy.
Such stars would be detected in the Galactic halo as hypervelocity
stars (HVSs) \citep{brown05}.  Assuming a threshold velocity of
$700\kms$ for escaping the Milky Way's potential, we find that about
15\% of all the escapers become hypervelocity stars. The average
velocity for the full sample of escapers is $\sim 350\kms$ for the
circular binary and $\sim 400\kms$ for the eccentric binary, while for
the sample of HVSs the mean is $\sim 1450\kms$ and $\sim 1550\kms$
respectively.  This is consistent with the theoretical estimate by
\citet{yt03}
\begin{eqnarray}
V_{\rm ej} & \simeq & \sqrt{\frac{3.2G\,\mbh\,\mim}{\left(\mtot\right) a}}\\
          & \sim & 2.3\times 10^3 \kms \left(\frac{\nu}{0.1}\right)^{1/2} 
\left(\frac{\mpc}{a}\right)^{1/2} \left(\frac{\mbh}{4\times10^6\msun}\right)^{1/2}
\end{eqnarray}
where $\nu = \frac{\mim}{\mtot}$.
The distribution of ejection velocities for escapers has a peak around
$\sim 100\kms$ and shows a long tail toward high velocities (up to
several thousand $\kms$).

\subsection{Tidal captures and mergers}
Interactions with the IMBH can also drive stars to be captured (either
tidally disrupted or accreted) by the massive black hole.  The
Schwarzschild radius for the Milky Way black hole is given by
\begin{equation}
  R_S = \frac{2 G \mbh}{c^2} \sim 3.8\times 10^{-7}\pc
\end{equation}
while the tidal radius for a solar type main-sequence star is
\begin{eqnarray}
R_t & = & R_\star \left(\frac{\eta^2 \mbh}{m_\star}\right)^{1/3} \\
 & & \sim 3.6\times 10^{-6} \pc  
\left(\frac{\mbh}{4\times10^6\msun} \right)^{1/3} 
\left(\frac{\msun}{m_\star}\right)^{1/3} 
\left(\frac{R_\star}{\rsun}\right)\,\eta^{2/3}\,.
\end{eqnarray}
For simplicity we set the capture radius for all stars to $R_t = 3
R_S$.  When a star approaches the SMBH within such distance, it is
merged with the black hole. The mass of the star is added to that of
the black hole and new positions and velocities are computed.  Mergers
occur when stars are scattered onto plunging orbits, and are therefore
rare events in the simulations.
\begin{figure*}
  \begin{center}
    \includegraphics[width=8.5cm]{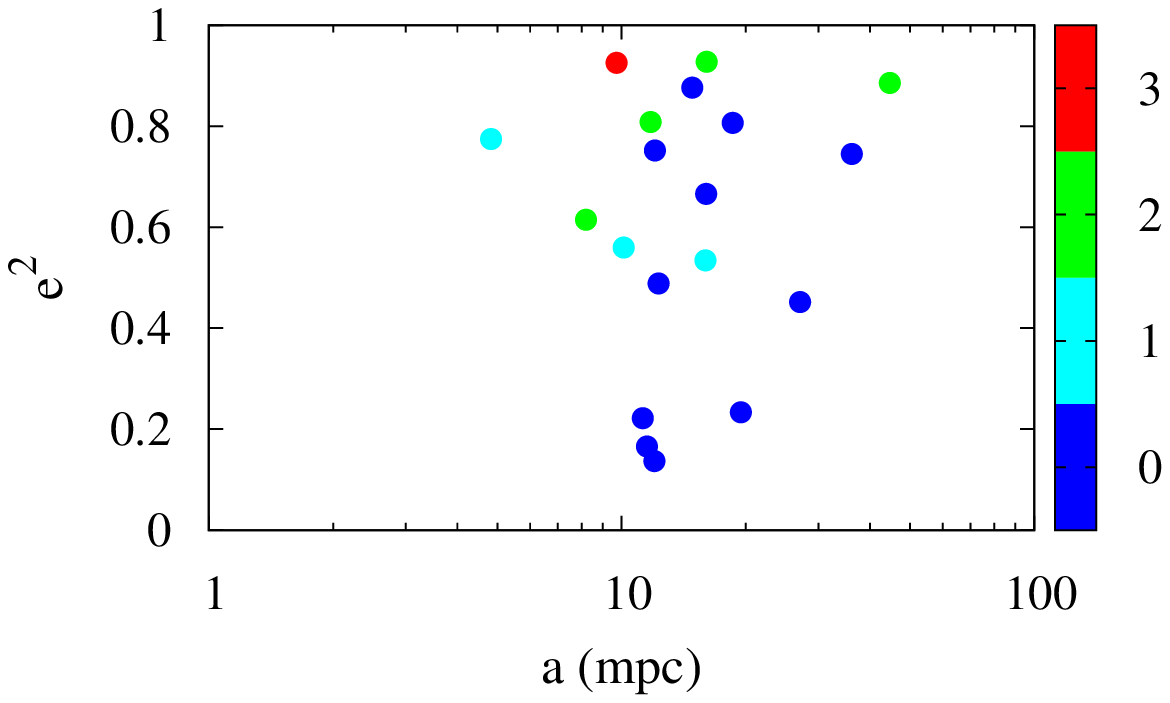}
    \includegraphics[width=8.5cm]{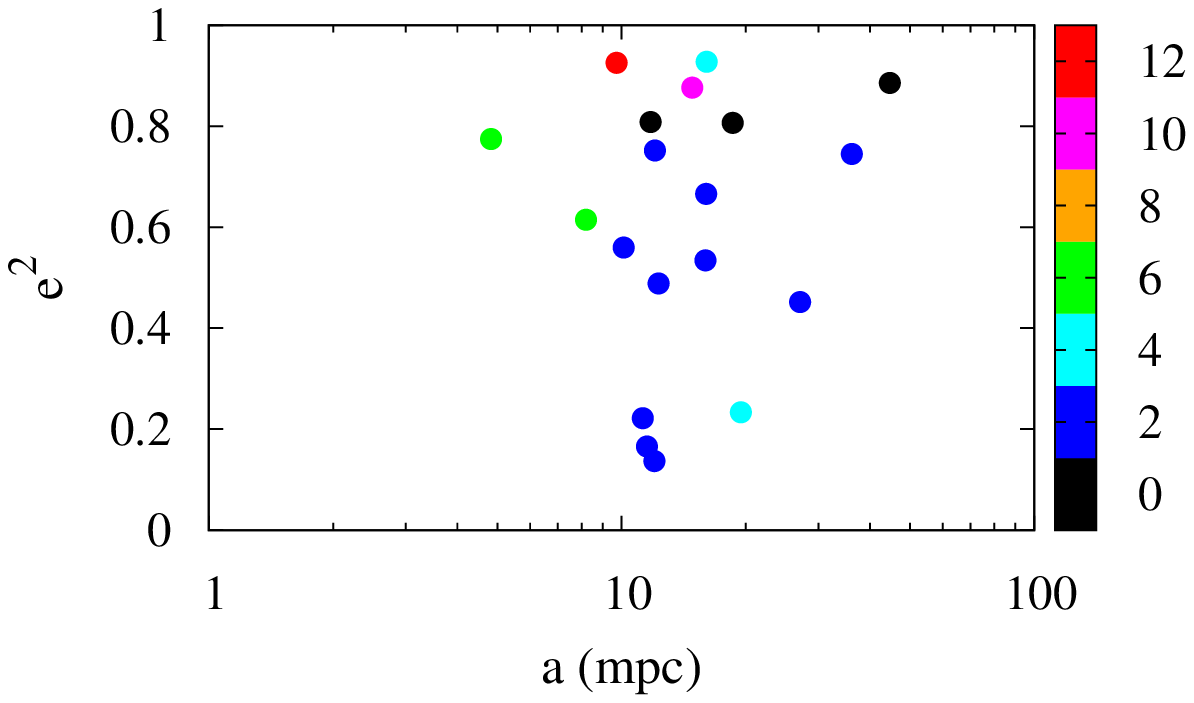}
  \end{center}
  \caption{Total number of mergers experienced by the individual stars
    as a function of their orbital parameters. Left: initially
    circular binary. Right: Initially eccentric binary.}
  \label{fig:mergers}
\end{figure*}
The total number of captures (out of 240 integrations) experienced by
the stars is 13 and 58 (corresponding to a probability of about 5 and
25\%) for the circular and eccentric case, respectively, indicating
that an eccentric binary is more likely to lead to a capture event.
Large eccentricity stars are more likely to be tidally disrupted by
the SMBH, as can be seen in Figure\,\ref{fig:mergers}.

Simulations with $q=10^{-3}$, $a=0.3\mpc$ and $e=0.7$ have the
shortest time for emission of gravitational waves, namely $\sim
2.7\myr$ (see Eq.\,\ref{eq:tgw}). For this choice of parameters, the
orbit of the IMBH decays due to energy loss, which is correctly
implemented in the code, until the two black holes coalesce.  From
this point on, the integration is carried on with only one black hole.

\subsection{Kozai oscillations}
The presence of an IMBH outside the orbits of the S-stars results in
perturbations on the inner stellar orbits.  In particular, Kozai
oscillations \citep{kozai62} can be induced in the stellar orbits if
the SMBH - star - IMBH system can be regarded as a hierarchical triple
with a large inclination between the inner and the outer orbit.  If
the relative inclination is large, the time-averaged tidal force
exerted on the inner binary produces variations in the eccentricity
$\einn$, inclination $j$ and argument of pericenter $\omega$. The
variations are such that the quantity
\begin{equation}
h = (1 -e^2)\, cos^2(j)\,,
\end{equation}
also known as {\it Kozai integral}, is conserved. This is proportional
to the component of the inner angular momentum perpendicular to the
outer orbit. The semi-major axis of the inner and outer orbits can
have periodic perturbations but no long-term evolution, so they can be
assumed to be roughly constant. While the evolution of the
eccentricity and inclination is always oscillatory, the argument of
pericenter may either oscillate ({\it libration}) or circulate ({\it
  circulation}). In either case the periodic oscillations in $\einn$
and $j$ are often called {\it Kozai cycles}. Remarkably, the
variations in $j$, $\einn$ and $\omega$ do not depend on the masses of
the stars or the dimensions of the orbits. These affect only the
timescale of the effect and not its size.

The characteristic time-scale for Kozai oscillations is given by
\citep{HHbook, KH07}
\begin{equation}
\label{eq:tkozai}
T_K = \frac{4 \mathcal{K}}{3\sqrt{6}\pi} \frac{\Pout^2}{\Pinn}
\frac{\mbh + \mstar}{\mim} \left(1-\eout^2\right)^{3/2}
\end{equation}
where $\Pinn$ and $\Pout$ are the period of the inner and outer
binary and $\mathcal{K}$ is a numerical coefficient which
depends only on the initial values of $j$, $\einn$ and $\omega$.

The maximum eccentricity $e_{\rm max}$ attained by the inner binary is
also a function of the initial values of $j$, $\einn$ and $\omega$
only.  In the case of a circular inner binary, $e_{\rm max}$ only
depends on the initial relative inclination. 

Kozai oscillations can be damped by general relativistic precession.

Three conditions must be satisfied in order for the Kozai mechanism to
operate in our simulations: (i) the IMBH orbit must lie outside the
orbits of the S-stars, so that the three body system can be regarded
as a hierarchical triple.  Whenever the inner and outer orbital
periods are well separated, Eq.~\ref{eq:tkozai} accurately predicts
the time scale for the oscillations in the eccentricity. The amplitude
of the oscillations can also be predicted from the initial parameters
\citep{KH07}.  (ii) The inner and outer orbits must be sufficiently
inclined with respect to each other. (iii) The period of the Kozai
oscillations must be shorter than the age of the system and shorter
than that of any other mechanism that might wash out the oscillations,
e.g. relativistic precession.

We therefore expect to observe Kozai oscillations in the simulations
with large $\aout/\ainn$ (like the runs with initial binary separation
$a=30\mpc$) for those stars whose orbits are significantly inclined
with respect to the SMBH/IMBH plane and with a long precession time
scale. Also, the timescale for oscillations will be shorter in the
runs with the largest IMBH mass ($q=10^{-3}$).

\begin{figure}
  \begin{center}
    \includegraphics[width=8.5cm]{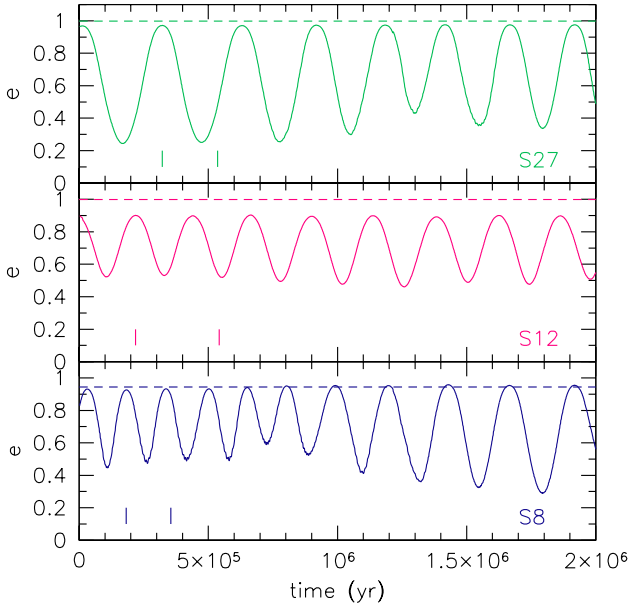}
  \end{center}
  \caption{Eccentricity evolution of stars S8, S12 and S27 in a run
    with $\mim = 4000\msun$, $a = 30\mpc$, $e=0$ and with relative
    inclinations of $j = 58^{\circ}$, $86^{\circ}$, $79^{\circ}$,
    respectively. Horizontal dashed lines represent the value of the
    maximum eccentricity predicted for the Kozai mechanism while the
    solid vertical lines indicate the Kozai timescale.}
  \label{fig:kozai}
\end{figure}
Figure\,\ref{fig:kozai} shows the evolution of the eccentricity for
three stars which exhibit Kozai oscillations in a run with $\mim =
4000\msun$, $a = 30\mpc$ and $e=0$. The predicted values of the
maximum eccentricity and of the Kozai timescale are also shown. The
agreement is satisfactory in all cases.  As expected, the semi-major
axis of the inner orbit remains constant and $h$ is conserved while
the eccentricity and the inclination oscillate.

\section{Effects on the observed population}
\label{sec:global}
The perturbations described in the previous section act
to modify the orbital elements of the stars.
\begin{figure}
  \begin{center}
    \includegraphics[width=8.5cm]{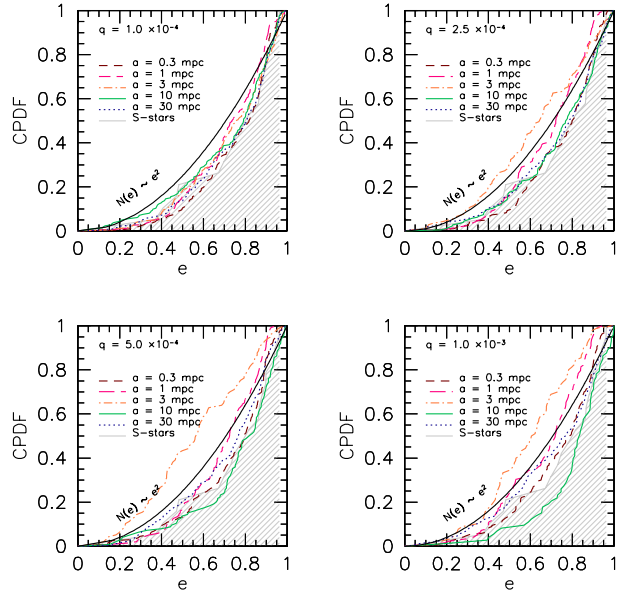}
  \end{center}
  \caption{Cumulative distribution function for the eccentricities of
    the S-stars after $10\myr$ of evolution in the case of a circular
    binary. Different lines are for different values of the binary
    semi-major axis. Each line is an average over the 12 different
    simulations with different IMBH 3D orientation with respect to the
    stars. The four panels refer to the four different values of the
    binary mass ratio. The shaded area represents the initial
    distribution of the stars. The $N(e) \sim e^2$ line corresponds to
    an isotropic distribution of eccentricities. }
  \label{fig:eMcirc}
\end{figure}
Figure\,\ref{fig:eMcirc} shows the cumulative distribution for the
eccentricities of the S-stars that remain bound at the end of the
simulations. The four panels refer to the four different values of the
binary mass ratio.  The distributions for $\mim=400-1000\msun$ are
consistent with an isotropic distribution of eccentricities $P(e) \sim
e$ for all values of the binary semi-major axis.  For larger black
hole masses ($\mim = 2000-4000\msun$), however, only very small
($a=0.3-1\mpc$) or very large ($a\geq30\mpc$) semi-major axes result
in final distributions consistent with observations.  Results from an
eccentric black hole binary appear qualitatively similar.  The figure
suggests that the IMBH exerts the largest perturbations at separations
of order $a\sim3-10\mpc$, which is comparable to the semi-major axes
of the S-stars. A close inspection of Figure\,\ref{fig:eMcirc} reveals
that the final distributions can deviate from the observed one in two
ways: either the final curve lies above the observed distribution, as
is the case for $a=1-3\mpc$, or it lies below the observed
distribution, as in the case of $a=10\mpc$. The reason for such
distinctively different behavior is the competition between two
effects of the IMBH perturbations: the tendency to increase the mean
eccentricity of the population and the removal of stars from the
population once they become unbound.  The former process tends to move
the distribution below the thermal line while the latter, by removing
the most eccentric stars, tends to push the distribution above the
thermal line. For semi-major axes $a=1-3\mpc$, the IMBH lies almost
completely within the orbits of the S-stars. Therefore, interactions
only occur with the stars that have small pericenter distances,
leading to ejections. This is consistent with the fact that the most
eccentric stars are the most likely to be ejected (see
Fig.\,\ref{fig:nef}) as well as with the fact that the largest number
of escapers is produced for $a=3\mpc$, regardless of the binary mass
ratio (see Fig.\,\ref{fig:enesc}). For $a=10\mpc$, the IMBH orbit
intersects those of all the S-stars, thus resulting in strong
perturbations on the orbital elements.  The global increase in the
eccentricities wins over the ejection of unbound stars in this case.
The relative importance of escapers over the global perturbations on
the orbital elements is also evident if we look at the mean
eccentricity of the sample for a given binary mass ratio and initial
separation. For $a=0.3$ and $a=30\mpc$, the value of the mean
eccentricity $\bar{e}$ remains close to that of the S-stars, which is
$\sim 0.75$. For $a=1-3\mpc$, when ejections dominate, the mean
eccentricity is reduced to $\sim 0.6-0.65$, while for $10\mpc$
$\bar{e}$ achieves the largest value of $\sim 0.8$.

\begin{figure}
  \begin{center}
    \includegraphics[width=8.5cm]{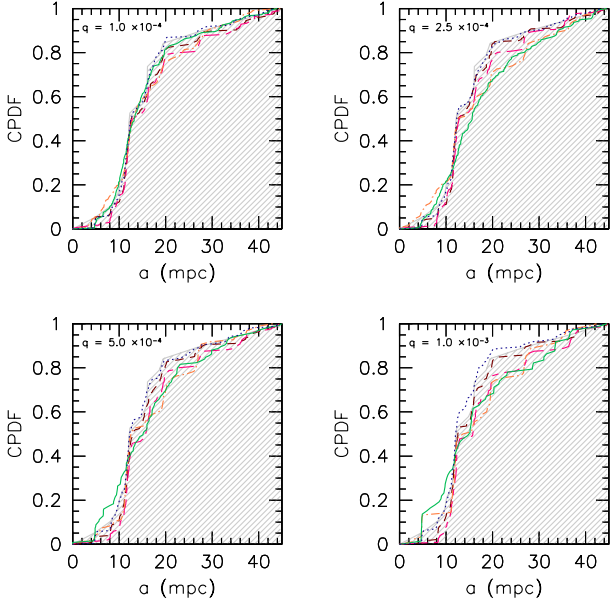}
  \end{center}
  \caption{Cumulative distribution function for the semi-major axes of
    the S-stars after $10\myr$ of evolution in the case of a circular
    binary. Line styles are as in Fig.\,\ref{fig:eMcirc}. }
  \label{fig:aMcirc}
\end{figure}
Figure\,\ref{fig:aMcirc} shows the cumulative distribution of the
semi-major axes of bound stars at the end of the simulations.  We
limit the range of semi-major axes to $45\mpc$, which is the maximum
value observed in the current S-stars sample. 
Deviations from the observed distribution occur preferentially for
large binary mass ratios and, in these cases, for binary separations
in the range $3-10\mpc$, similarly to what we observe for the
eccentricity distributions.

Both Figure\,\ref{fig:eMcirc} and \ref{fig:aMcirc} show distributions
after $10\myr$ of integration. The evolution in the distributions,
however, is expected to be time-dependent since the perturbations from
the IMBH and the ejections tend to grow in time (see
Fig.\,\ref{fig:tmean}). We find that the effect of ejections becomes
apparent after only $1-2\myr$ and after $5\myr$ the distributions take
the form they have at the end of the simulations.

\section{Discussion}
Our results help place limits on the orbital parameters of possible
IMBHs in the Galactic center.
Constraints have been placed in the past by several authors.
One natural constraint comes from the requirement that the center of mass
of the binary coincides with the peak of the stellar distribution 
within the observational uncertainties \citep{yt03}
\begin{equation}
\frac{\mim}{\mim+\mbh} \abh \simless 8\mpc\,,
\end{equation}
where $\abh$ represents the binary semi-major axis.
\begin{figure}
  \begin{center}
    \includegraphics[width=9.0cm]{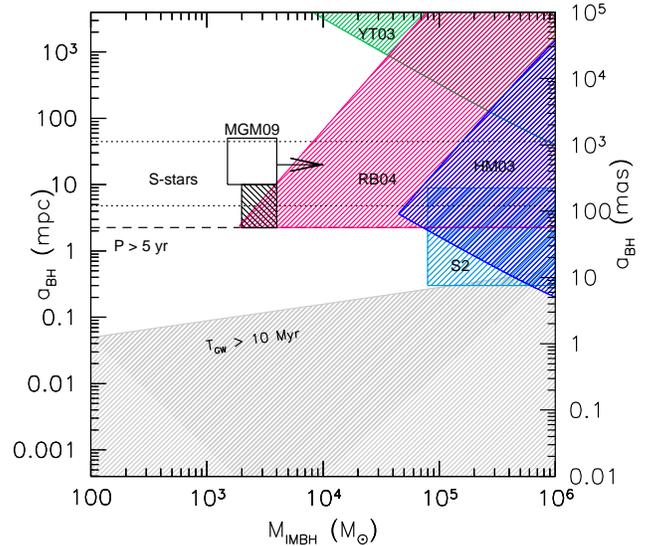}
  \end{center}
  \caption{Constraints on the orbital parameters of a hypothetical
    IMBH in the Galactic region. The shaded areas represent regions of
    parameter space that can be excluded based on observational or
    theoretical arguments. The dotted lines mark the distances at
    which the S-stars are currently observed. The dashed line
    represents the $5\yr$ orbital period corresponding to discoverable
    systems. The parameters enclosed in the empty rectangular box are
    required for an efficient randomization of inclinations in the
    cluster infall scenario \citep{mgm09}. The small rectangular
    region just below the empty box represents the parameter space
    excluded by this work.}
  \label{fig:final}
\end{figure}
The corresponding prohibited region in the $\mim - \abh$ plane is
shown in the upper right corner of Figure\,\ref{fig:final}.

The orbit of a black hole binary shrinks due to emission of
gravitational waves on a timescale given by Eq.\,\ref{eq:tgw}.
Demanding a lifetime of at least $10\myr$ excludes the lowest region
in Fig.\,\ref{fig:final} corresponding to small separations.

\citet{gill09} derive an upper limit of $0.02\,\mbh$ to the mass that
can be contained within the orbit of star S2. This implies that no
more than $8\times10^4\msun$ can be hidden between $\sim 0.3\mpc$ and
$\sim 9\mpc$ of the SMBH.

The most stringent constraints can be derived from the astrometric
wobble of the radio source SgrA*, which could show indication of
binary motion. Very Long Baseline Array observations from 1995 to 2000
show a residual motion for SgrA* of $\sim 0.5\mas = 0.02\mpc$ and an
upper limit of $\sim 8\kms$ for the peculiar motion perpendicular to
the Galactic plane. Such limits exclude the rightmost triangular region
in Fig.\,\ref{fig:final}) \citep{hm03, yt03}.  More recent
measurements from \citet{rb04} find an upper limit to the motion in
Galactic latitude of $-0.4 \pm 0.9 \kms$. Combining this restriction
with the $5\yr$ limit on observable systems results in the large
triangle region in Fig.\,\ref{fig:final}.

The perturbations exerted by an IMBH on the orbits of the S-stars
allow us to exclude the region of parameter space corresponding to
masses $2000-4000\msun$ and initial semi-major axes $\sim 2-10\mpc$.
Such region is represented by the shaded box in the figure.

Interestingly, the IMBH parameters required for an efficient
randomization of inclinations \citep{mgm09} in the cluster infall
scenario ($\mim \simgreat 1500\msun$ for the simulated range of
separations $10-50\mpc$; see rectangular box in Fig.\,\ref{fig:final})
are consistent with all the constraints placed so far.  Mass ratios
larger than $10^{-3}$ were not simulated but they are expected to
operate even faster than the smaller ones.

\acknowledgments
We thank Douglas Heggie and Hagai Perets for useful discussions.  This
work was supported by grants AST-0206031, AST-0420920 and AST-0437519
from the NSF, grant NNG04GJ48G from NASA, and grant HST-AR-09519.01-A
from STScI.

\end{document}